\documentclass[acmlarge]{acmart}

\usepackage{xcolor}

\AtBeginDocument{
  \color{black}
}

\AtBeginDocument{%
  }

\usepackage{booktabs}
\usepackage{longtable}
\usepackage{array}
\usepackage{listings}
\usepackage{pifont}

\setcopyright{cc}
\setcctype{by}
\acmJournal{IMWUT}
\acmYear{2026} \acmVolume{10} \acmNumber{4} 
\acmMonth{12} 
\begin{document}

\title{When Data Becomes Judgment: Misaligned Interpretations and Accountability in Food Delivery Platforms}

\author{Yibo Meng}
\orcid{0009-0009-0370-5456}
\affiliation{
  \institution{Cornell University}
  \city{Ithaca, NY}
  \country{United States}}
\email{yim4007@med.cornell.edu}

\author{Shuoning Shi}
\orcid{0009-0003-2345-6449}
\affiliation{
  \institution{University of Illinois Urbana-Champaign}
  \city{Champaign, IL}
  \country{United States}}
\email{sshi14@illinois.edu}

\author{Bingyi Liu}
\orcid{0009-0000-4312-153X}
\affiliation{
  \institution{University of Michigan}
  \city{Ann Arbor, MI}
  \country{United States}}
\email{bingyi@umich.edu}

\author{Hongkai Chen}
\authornote{Corresponding author.}
\orcid{0000-0001-7206-6584}
\affiliation{
  \institution{The Chinese University of Hong Kong}
  \city{Hong Kong}
  \country{China}}
\email{hkchen@ie.cuhk.edu.hk}

\begin{abstract}
Food delivery platforms mediate service encounters through real-time tracking data. 
While presented as objective, such data often obscures the situational constraints shaping delivery work, producing systematic misalignments between user perception and courier experience.
This study presents a socio-technical analysis of real-time mobile tracking systems in the wild. 
Through semi-structured interviews with 23 users and 17 couriers on Chinese food delivery platforms, we identify two interrelated dynamics. 
First, users operate within a data-as-behavior interpretive framework, translating spatial and temporal anomalies into moralized judgments of courier negligence.
Second, couriers engage in anticipatory data management, a form of hidden digital labor in which they reshape their physical behavior to produce interface-legible trajectories rather than physically optimal ones.
Together, these findings expose a burden-shifting mechanism---characterizing the systemic outcomes of decontextualized interface design rather than explicit designer intent---in current tracking architectures, demonstrating how current tracking architectures leave gig workers bearing much of the explanatory burden.
We propose design directions toward contextual transparency, redistributing this explanatory burden from individual workers to the platforms that possess the logistical context to bear it.
\end{abstract}


 
\maketitle

\section{Introduction}
\label{introduction}

Food delivery platforms are deeply embedded in urban life, mediating routine interactions among users, couriers, and restaurants on a massive scale. {In China alone, platforms like Meituan processed over 150 million daily orders in July 2025 ~\cite{meituan_q2_2025}.}
As ubiquitous computing infrastructures, these systems transform physical movement into a continuous stream of real-time sensing data, such as GPS locations, estimated times of arrival (ETAs), and stage-based status updates, presented via mobile interfaces.
While Ubicomp and infrastructure studies traditionally
view visibility and awareness as  essential resources for coordination~\cite{Dourish1992,Schmidt2002}, treating tracking systems merely as neutral informational layers obscures how  their technical design reshapes worker behavior.
Existing research highlights that delivery work is organized through algorithmic management, using data-driven evaluation to dictate schedules, movements, and accountability~\cite{huang2023algorithmic}.

This paper contributes
to the Ubicomp community by offering a socio-technical analysis of real-time mobile tracking systems in the wild.
{We argue these sensing interfaces are not merely transparent informational channels; rather, they act as diagnostic sites revealing how technical systems fail to communicate situated context.}
Greater visibility has not universally smoothed coordination, and tensions between users and couriers remain pervasive. 
Tracking interfaces typically present platform data as objective, self-evident traces, providing minimal access to the complex realities of delivery work, such as traffic, batching, route adjustments, restaurant delays, and access constraints~\cite{huang2023algorithmic,Shaikh2023,Xu2025}. 
This creates a \textit{transparency paradox}: users see high-granularity spatial data, but often understand less of the environmental conditions behind it. 
Crucially, this informational asymmetry is unevenly distributed. Users bear the interpretive burden of making sense of decontextualized traces, while couriers shoulder the relational and reputational fallout of the resulting judgments.

While prior research extensively examines platform labor as a site of algorithmic control, the interpretive role of user-facing data within the user-courier dyad remains underexamined~\cite{Chen2022}. 
Existing studies primarily view platform data as tools for efficiency, coordination, or managerial control, with adjacent research noting how workers adapt via anticipatory compliance and impression-management strategies~\cite{Bucher2021}. 
However, we know significantly less about how users moralize ostensibly neutral tracking traces, translating a stationary GPS dot into laziness or an extending ETA into a broken promise, and how couriers, in turn, reshape their physical labor in anticipation of this evaluative gaze.
Moving beyond traditional labor studies, we focus on the ``data-as-behavior'' feedback loop: how real-time system outputs (sensing accuracy, update frequency, and ETA modeling) directly dictate the physical decision-making of the couriers.
Through a qualitative study of 40 participants (23 food delivery users and 17 couriers), we explored the social life of delivery data via semi-structured interviews.
By juxtaposing how both groups perceive the same interface data, specifically GPS locations, ETAs, and status nodes, we reveal how mobile-sensing outputs become moralized sites for accountability.
Our research was guided by the following research questions:
\begin{itemize}
    \item RQ1: How do users interpret platform data (e.g., location, time, routes) and translate it into judgments of courier behavior and accountability?
    
    \item RQ2: How do couriers perceive user-facing data presentations, and how does this awareness shape their decision-making and actions during delivery?
    
    \item RQ3: What are the expectations of both users and couriers for future computing systems to resolve these interpretive misalignments?
\end{itemize}

Across them, we identified a profound misalignment between data representation and delivery practice. 
{Users frequently treat spatial and temporal anomalies as evidence of courier intent.
Aware of being judged through these interface traces, couriers respond by strategically adjusting their physical trajectories to maintain interface legibility.
This demonstrates that real-time tracking does not merely communicate progress; it actively shapes judgment, accountability, and courier behavior in unintended ways.}
The main contributions of this paper are summarized as follows:

\begin{itemize}
    \item We provide a dual-perspective empirical analysis of real-time tracking systems in the wild, grounded in semi-structured interviews with 23 users and 17 couriers, to juxtapose how the same interface data is experienced from distinctly different practical positions.
    
    \item  We empirically characterize a data-as-behavior interpretive framework among users, showing how apparently objective tracking metrics are translated into moralized judgments of individual negligence. 
    
    \item  We identify anticipatory data management as a strategy where couriers prioritize the appearance of their delivery trajectory over physical efficiency to mitigate complaint risks and manage the data shadow.
    
    \item We conceptualize a burden-shifting mechanism inherent in current tracking architectures, showing how these architectures leave gig workers performing much of the cognitive and emotional labor of explaining systemic logistical failures.
    
    \item  We propose 
    design directions for contextual transparency that redistribute explanatory labor from individual workers to platform infrastructures, ensuring accountability aligns with the true logistical context of the work.
\end{itemize}

\section{Related Work}

{The Ubicomp community has increasingly transitioned from studying interpersonal location sharing to conducting socio-technical examinations of platform systems that orchestrate human labor and social relations. 
Our work bridges four key domains: location sharing, algorithmic management, infrastructure studies, and measure-target shift.

\subsection{Location Sharing: From Social Coordination to Asymmetric Purpose-Driven Systems}

The Ubicomp community has long studied location sharing, initially framing it around interpersonal social awareness and coordination. 
Early research
explored how location data could support family values and negotiate daily routines~\cite{brown2007locating},
and facilitate spontaneous social connection rather than rigid surveillance~\cite{humphreys2007mobile,barkhuus2008awareness}.
These foundational studies established that peer-to-peer location sharing is fundamentally oriented toward fostering social awareness and communication among trusted groups, inevitably linking perceived utility to social relationship.

As these systems proliferated, research expanded to the complexity of privacy and user consent.
Privacy management is rarely static; it is highly context-dependent and relies heavily on the perceived closeness between sharer and viewer~\cite{wiese2011close}. 
Early frameworks embedded privacy policies in mobile  networks~\cite{sadeh2009understanding}, while later work empirically modeled the trade-offs between sharing benefits and privacy costs~\cite{toch2010empirical}.
Crucially, privacy in location sharing is adaptive:
Wilson et al.~\cite{wilson2013privacy} demonstrated that users actively manipulate their sharing behaviors as they acclimate to an application's utility and risk. 
Tang et al.~\cite{tang2010rethinking} further articulated a critical distinction between ``social-driven'' sharing (maintaining social bonds) and ``purpose-driven'' sharing (utility and logistics).

Food delivery platforms represent a contemporary evolution of ``purpose-driven'' sharing. 
Unlike the voluntary, peer-to-peer nature of classic Ubicomp applications, location sharing in the gig economy is compelled as a condition of service within a fundamentally asymmetric, transactional relationship. 
This context introduces new frictions: ``sharing'' is no longer about mutual benefit, but about algorithmic accountability and the moralized interpretation of spatial behavior. 
Our study extends this literature by examining purpose-driven sharing environments where location transparency is actively used for courier behavior management and customer judgment.

\subsection{{Algorithmic Management: Information Asymmetry and Worker Adaptation}}

Algorithmic management refers to the use of automated track-and-trace technologies and data analytics to perform functions traditionally handled by human managers~\cite{jarrahi2021algorithmic}. 
While early Ubicomp research focused on optimizing algorithm efficiency, such as objective performance modeling~\cite{mirjafari2019differentiating}, urban GPS improvement~\cite{liu2018foodnet}, and ETA prediction refinement~\cite{wen2023enough},
recent attention has shifted toward the human cost of these optimizations~\cite{russo2023urban,baseman2025clinical}.

Delivery platforms orchestrate courier work through data infrastructures, including dispatch logic, ETA calculations, and completion-time targets~\cite{griesbach2019algorithmic,huang2023algorithmic,tuomi2024strategies}. 
These systems translate delivery work into measurable indicators that shape couriers' schedules, movement rhythms, and working conditions~\cite{duggan2023algorithmic,DongZhangWu2025Burnout}. 
Furthermore, ranking systems, punitive mechanisms, monitoring practices, workflow design, customer feedback, and restaurant delays can significantly contribute to courier burnout~\cite{DongZhangWu2025Burnout}. 
{On the user side, the rating system functions as a governance structure, where complaints and satisfaction metrics directly impact couriers’ income, future order allocation, and occupational stability~\cite{Rosenblat2017,Hernandez2024}. 
However, less is known about how users form these judgments based on interface metrics like maps, ETAs, route traces, and status updates.
Because courier labor is organized through concealed platform algorithms~\cite{Lee2015},
this management style creates a severe information asymmetry: platforms possess vast data on courier behavior but provide workers with minimal, actionable instructions~\cite{he2021demand}.}

In response, couriers develop adaptive strategies, such as selectively accepting orders and dynamically adjusting routes~\cite{Kusk2022,Chen2022}.
Courier decision-making is heavily influenced by psychological perceptions of distance and deadlines rather than strict distance optimization~\cite{zhang2019route},
creating a ``calculative landscape''~\cite{lu2026ubiquitous,sun2019your}, where couriers engage in ``algorithmic play'' or workarounds to maintain performance metrics.
This gig labor requires continuous interpretive effort to anticipate opaque platform logics~\cite{Shaikh2023}, while the work's individualized structure produces atomization, forcing workers to selectively exchange information while competing for scarce opportunities~\cite{Yao2021}. 
Thus, couriers manage both their deliveries and their ongoing relationship with the system.
Our study contributes to this domain by identifying an information architecture in which system-wide context is not surfaced to workers, a structural condition that leaves couriers to manage the resulting social friction.

\subsection{Infrastructure Studies: Visibility Without Causal Intelligibility}

Visibility has long been treated as a coordination resource that reduces uncertainty through shared awareness~\cite{kwon2018connected,Dourish1992,Schmidt2002,kang2025you,weinshel2025would}.
While delivery platforms extend this logic via real-time tracking cues (e.g., live maps and ETA countdowns)~\cite{Xie2024},
these interfaces offer visibility without causation, displaying processed outputs rather than the operational conditions that generate them~\cite{RosenblatStark2016,Cheon2025}. 
Users see spatial and temporal courier data but lack access to the underlying context, such as restaurant delays or dispatch logic. Our theoretical framework draws on Star and Ruhleder's conceptualization of infrastructure as a system that ``sinks into the background'' and becomes visible primarily through failure~\cite{star1994steps}. 
While traditional Ubicomp research often champions seamless experiences, revealing system seams can empower users to understand and navigate inherent limitations~\cite{chalmers2004seamful}.
Our findings suggest that many operational seams remain invisible in user-facing delivery interfaces. This limited visibility leaves platform-level causes largely unavailable for accountability, requiring couriers to explain operational disruptions to users.

This invisibility is particularly problematic regarding temporal uncertainty~\cite{preyss2007stochastic}. 
Studies on bus arrival times highlight a critical tension: point predictions (e.g., ``Arriving in 10 minutes'') offer comforting certainty but create false precision that fails to account for stochasticity~\cite{kay2016ish}. 
Food delivery platforms exacerbate this by bundling systemic delays  with individual courier performance.
When a point ETA prediction is violated, users rarely attribute the delay to systemic variables like restaurant backlogs. 
Instead, the algorithm's uncertainty materializes as a failure of the individual courier.
By presenting high-granularity data without situational interpretability, the interface redistributes uncertainty: users are left to infer causes from incomplete information, while couriers bear the reputational consequences.

\subsection{The Measure-Target Shift: From Quantitative Metrics to Social Judgment}

{In the gig economy, quantitative metrics like ETAs often undergo a ``measure-to-target'' shift, a classic manifestation of Goodhart's Law~\cite{goodhart2015goodhart}.
When a metric functions simultaneously as a rigid performance target for couriers and a service promise for users, it ceases to be a reliable measure of quality.
This structural misalignment is reinforced by interface design: metrics are presented to users as binding commitments, while the underlying logistical context that might explain deviations is often absent or only weakly represented in the user-facing interface.}

Research suggests that when encountering outcomes without sufficient contextual information, individuals often attribute those outcomes to personal traits (e.g., effort or competence) while underweighting situational constraints~\cite{GilbertMalone1995,JonesHarris1967,HanLiuLoewenstein2023}. 
{Consequently, interface traces intended to represent objective system states, such as stalled movement or route deviations, are routinely repurposed as evaluative cues~\cite{Ravula2023DeliveryPerformance}.
Tracking data invites moralized inferences about a courier's reliability~\cite{Quarles2025,DongZhangWu2025Burnout},

Furthermore, our analysis shows that platforms do not merely transmit information; the resulting interface configurations distribute interpretive labor and accountability in systematic ways, which directly affects labor conditions through ratings, complaints, and negative reviews~\cite{Zong2024,Hsieh2023}.
To avoid penalties, couriers adapt behaviors like route shape and pace specifically to manage the customer's perception~\cite{DengTangLai2026,Xu2025}. 
This creates a recursive loop:  interface ambiguity drives user inference, which in turn forces worker adaptation~\cite{Sekharan2025}.}

{By converting shared logistical challenges into individualized moral judgments, the resulting interface structure can make platform-level processes less visible while leaving couriers and users to bear much of the interpretive labor. 
Our work bridges these disparate perspectives, demonstrating how the platform's infrastructure converts logistical uncertainty into a localized conflict between two end-users.}

\section{Methodology}

In this section, we describe the methodology used in this paper to conduct qualitative studies.

\subsection{Research Design}

This study adopts a qualitative research design using semi-structured interviews to explore how data presentation in food delivery platforms is understood by users and couriers, and how it subsequently shapes their behaviors and responsibility attributions.  
By focusing on participants' concrete experiences with spatial and temporal data, we juxtapose the perspectives of both roles to capture interpretive differences stemming from their distinct informational and practical positions.
Analyzing these experiential narratives reveals how computing systems translate interface data into behavioral meanings and moralized judgements in everyday practice.

\subsection{Participants Recruitment and Demographics}

We recruited 40 participants, including 23 users (U1--U23) and 17 couriers (F1--F17), via Chinese social media platforms (WeChat, RedNote, Baidu Tieba, Douyin, Weibo).
User participants (age 19--55; 10 urban, 13 rural) spanned diverse occupations and educational backgrounds.
Courier participants (ages 24--45; 7 female, 10 male; 11 urban, 6 rural) similarly reflected varied educational levels. 
See Supplementary Table 1 for demographic details.
We sought participants with at least six months of platform experience, ensuring a diverse range of usage frequencies and life rhythms without rigid screening criteria. 
Crucially, these two groups operate under vastly different informational conditions: users interpret delivery progress via interface traces, whereas couriers navigate both system data and real-world constraints. This structural asymmetry provided a multidimensional basis for analyzing data interpretation and behavioral adaptation.


Specifically, the recruitment information briefly described the research topic (focusing on data understanding and usage experiences in the food delivery process), participation format (semi-structured interviews), expected duration, and basic participation requirements (at least six months of experience in using or delivering food delivery services).  
After registration, the researchers conducted brief communication to understand participants' basic conditions, including frequency of food delivery use, usage scenarios, and general life rhythms.  
Without setting strict screening criteria, efforts were made to ensure diversity in usage experience.

\subsection{Research Ethics}

The study was approved by our institutional ethics review board.
Participants provided informed consent after receiving clear explanations of the research purpose, data usage, and their right to withdraw. 
All data were anonymized during transcription, using identifiers (e.g., U1, F1).
Interviews were conducted following the principle of minimal risk; researchers actively avoided leading questions and terminated sensitive topics to minimize psychological burden or real-world occupational risk.
Data is stored on encrypted devices with access strictly restricted. 

\subsection{Data Collection Procedures}

We conducted semi-structured interviews (45--65 minutes, audio-recorded with consent) tailored to the participants' respective roles within the delivery ecosystem.

\subsubsection{User Interviews}

User interviews primarily focused on information perception, interpretation processes, and judgment formation.  
At the outset, researchers introduced the study's purpose, emphasizing a focus on real-world ordering experiences rather than platform evaluation.  
Open-ended questions then guided participants to recall their typical usage routines, such as describing ``a complete experience from placing an order to receiving the food,'' to help them ground their reflections in concrete situations.  

During the subsequent experience elaboration stage, participants recalled one or two memorable deliveries, encompassing smooth, abnormal, or confusing scenarios.  
Follow-up questions, such as ``what happened'' and ``how did you understand it at the time,'' encouraged detailed descriptions of these specific events.  

The interview then transitioned to the key mechanism exploration stage.
Researchers asked questions like, ``when would you feel that something is not quite right with this order?'' to identify critical moments of judgment during the waiting process.
To prevent priming, we deliberately avoided using predefined data dimensions (e.g., explicitly asking about location, time, or routes) as a questioning framework.
Instead, participants were encouraged to describe the information they naturally attended to.
We also asked how their interpretive approach had evolved over time and what specific records they would want during a delivery dispute.

Once key situations were identified, researchers probed the judgment formation process with questions such as, ``how did you make that judgment,'' ``what information did you rely on,'' and ``why did this information lead you to such an understanding.'' 
This process revealed how users translate interface cues into status assessments and subsequent responsibility judgments.
Simultaneously, we asked their behavioral responses to uncertainty, e.g., whether they chose to wait, repeatedly check the interface, or actively contact the couriers.

The interview further explored experiences of a ``mismatch between data and reality,'' asking participants to recall instances where interface displays contradicted actual conditions and how they retrospectively comprehend those discrepancies.
In the final stage, discussions shifted to user expectations for future systems. 
Participants were asked ``what kind of system could reduce uncertainty,'' ``what information could help better understand the process,'' and ``what kind of records or evidence you would expect when problems occur'' (see Supplementary Table 2).


\subsubsection{Courier Interviews}
Courier interviews centered around action decision-making, data understanding, and the anticipation of user perspectives during deliveries.
Following an introduction to the study's purpose, participants were asked to ground their accounts in their everyday delivery work by describing their typical workflow, framing it around ``the general process from accepting an order to completing delivery.''

During the experience elaboration stage, participants recalled one or two memorable deliveries, including both smooth and difficult scenarios. 
Follow-up questions extracted the details of these events, aiming to capture concrete practical experiences in real working environments.

The interview then moved to the key situation identification stage. 
Researchers asked questions such as “when would you feel that this order might cause problems” or “when would you feel that users might have complaints” to help participants pinpoint moments of active risk judgment. 
Participants were also asked about moments of feeling understood or misunderstood by customers, and how their judgment had changed with experience.

Upon identifying these critical situations, researchers explored the underlying bases for their judgments and subsequent action adjustments with questions like ``how did you make that judgment'' and ``why did you handle it that way.''
This step revealed how couriers combine platform data with real-world conditions to understand situations and make decisions during deliveries. 
We then explored couriers' experiences of a ``mismatch between data and reality'', instances where interface displays diverged from physical conditions, and analyzed how these discrepancies impacted their work strategies and customer interactions.


Finally, the interview concluded by discussing couriers' expectations for future systems, specifically focusing on ``how to reduce misunderstandings,'' ``what information should be presented by the system,'' and ``how to reduce the burden of explanation'' (see Supplementary Table 3). 

\subsection{Thematic Analysis and Mechanism Abstraction}

We employed thematic analysis combined with mechanism-oriented abstraction to extract how users and couriers form judgments and respond behaviorally.
Given the significant differences in practical position, information visibility, and action goals, the user and courier datasets were coded and assessed for saturation independently, and cross-role comparative analysis was conducted on this basis.

During data preparation, all interviews were audio-recorded with participant consent and transcribed verbatim.
To faithfully preserve participants' original meanings, the transcriptions retained pauses, repeated expressions, and emotional cues. 
The research team then conducted multiple read-throughs to familiarize themselves with the dataset, recording preliminary observations without imposing predefined categories.

Initial coding was performed separately for the user and courier datasets.
Four researchers selected a subset of interviews (approximately 3--4 per group) and conducted line-by-line coding to identify meaningful expression segments. 
To accurately capture participants' situated understandings, judgments, and actions, the codes remained close to the original expressions.
The research team subsequently discussed these initial results to align coding granularity, naming conventions, and boundaries, eventually developing a shared codebook.
The remaining data were divided among the researchers; each transcript was analyzed by a primary coder and verified by a secondary reviewer.
Discrepancies were resolved through regular meetings and by referencing the original transcripts until consensus was reached.

Category development followed the coding phase.
Rather than grouping codes by data type, they were organized by their functional roles in cognition and action.
For example, varying types of information were considered as ``bases for judgment,'' while various anomalous experiences were grouped as ``situations triggering interpretation.'' 
This analytical shift from focusing on what the information is to what the information does allowed us to form intermediate categories with explanatory power.
Building on this, we conducted mechanism abstraction.
The research team reorganized categories to uncover underlying shared logics and functional pathways.
For user data, the analysis focused on how interface cues translate into process comprehension and responsibility judgments.
For courier data, the analysis centered on how user-visible data are internalized as action constraints that guide anticipatory adjustments. 
Through this process, we synthesized a series of mechanism-based explanations.  

We assessed data saturation separately for the two groups. 
For the user dataset, theoretical saturation was reached after the 18th participant, as no new interpretive paths or judgment patterns emerged.
Similarly, the courier dataset reached saturation after the 14th participant, with no new behavioral strategies or cognitive patterns identified. 
Subsequent interviews validated the stability of the existing categories and mechanisms without introducing novel core structures. 

Following these independent analyses, we conducted a cross-role comparison. 
By comparing how users and couriers interpreted identical data phenomena, we identified systematic differences in meaning construction and responsibility attribution.
For instance, while users frequently translated data anomalies directly into behavioral flaws, couriers interpreted the same anomalies as environmental or systemic constraints. 
This comparison highlighted the differentiated roles data play across distinct practical positions.
Additionally, our analysis continuously accounted for negative cases, i.e., individual experiences that deviated from dominant patterns.
For example, some users actively questioned the platform's data reliability, while certain couriers did not alter their behavior based on user-facing metrics.  
Analyzing these outliers helped rigorously define the boundaries of our mechanisms and prevented overgeneralization.

Overall, through role-separated coding, independent saturation assessments, and collaborative abstraction, this study constructs mechanism-based explanations grounded in real experiences, yielding a systematic analytical foundation for understanding how platform data functions in practice.

\section{Empirical Findings: The Socio-Technical Life of Tracking Data}

In this section, we describe the analytical results of our semi-structured interviews.

\subsection{From Spatial Traces to Moralized Attribution: How Users Construct Responsibility}

Our qualitative analysis reveals that users systematically convert raw interface metrics into moralized attributions of worker capability and intent. 
Across daily ordering experiences, users rely on four primary informational dimensions, i.e., location movement, estimated arrival times, route trajectories, and staged status nodes, to construct responsibility judgments when deliveries deviate from expectations.

\subsubsection{Location and Movement Status as Behavioral Proxies}

Users routinely treat a courier’s location and movement status as direct behavioral evidence, interpreting prolonged stationary periods or slow movement as intentional delay or inefficiency while overlooking logistical realities such as food preparation delays, building access barriers, or signal blackouts. 
This interpretive tendency is conditioned by interface design: continuous trajectories and near real-time updates present heavily processed algorithmic data as transparent spatial reality.
As one courier, F2, observed, users \textit{``cannot see things like system rules, order dispatching, or the restaurant’s meal preparation rhythm,''} leaving them to fixate entirely on the binary of \textit{``movement and non-movement on the map.''}
Because platform data is inherently ``mimetic''----presenting polished representations that mask underlying operational discrepancies, as F6 noted---users readily mistake interface traces for objective facts. 
Consequently, when signal drops or coverage gaps occur, users attribute connectivity failures to personal negligence, as U11 experienced when mistakenly criticizing a courier passing through a zone with \textit{``poor network connectivity.''}

When tracking indicators stops, users immediately convert interface immobility into attributions of bad faith. 
Confronted with a static icon, users like U4 expressed incredulity---asking \textit{``how could they possibly not move at all?''}---while relying on the assumption that a \textit{``real-time''} map guarantees that a lack of interface updates means the worker is \textit{``not been doing anything.''} 
This reading reduces complex logistical friction to personal misconduct; observing \textit{``just one dot sitting there without moving,''} U5 described feeling that couriers were deliberately \textit{``dragging out the time or busy with another order,''} emphasizing that \textit{``if they were waiting for the food or something, the platform should say so.''}
Absent contextual explanations from the interface, users fill the informational void by moralizing spatial stasis into perceived worker accountability.

\subsubsection{Algorithmic ETAs as Binding Commitments and Reference Anchors}

Because users lack insight into how estimated delivery times are generated, they often treat algorithmic ETA predictions as reliable, binding commitments rather than dynamic estimations. 
Consequently, users rely on the estimated arrival time as a rigid reference to organize their daily schedules and personal activities. 
When delivery fails to align with this expectation, users immediately attribute the discrepancy to courier negligence. 
This dynamic is further intensified by local accountability mechanisms---such as platform policies where customer complaints trigger default fault attribution and financial penalties for workers~\cite{Chen2022}---giving ETA deviations severe consequences. 

Because the platform presents predictions as precise minute counts and countdowns, dynamic multi-variable calculations are compressed into perceived delivery guarantees. 
As U1 argued, once \textit{``the delivery time point [is] clear,''} users expect couriers to \textit{``abide by this time node, rather than breaching it,''} questioning \textit{``what would be the point of having this time node at all?''}
Similarly, U13 explained organizing personal routines like \textit{``rest times''} around the platform's schedule, maintaining that couriers failing to meet this deadline should face penalties for \textit{``breach of contract.''} 
When these fixed temporal expectations are violated, user frustration is directly funneled into judgments of personal blame. 
Reflecting on habitual ordering patterns, U9 expressed immediate distress when a newly assigned courier was \textit{``late on the very first day,''} immediately treating the delay as a personal failure. 
Drawing from operational experience, F12 highlighted how platform ETAs exacerbate this attribution bias, noting that platforms recklessly present predictions that cause \textit{``a great deal of negative impact''} because users \textit{``really take that time as reality,''} automatically assuming that \textit{``all the responsibility lies with the courier not being conscientious enough.''}

\subsubsection{Route Deviations as Indicators of Opportunistic Detours}

Users naturally evaluate delivery routes through the lens of everyday travel, where the most direct, shortest path is presumed to be the sole indicator of efficiency. 
When a courier's trajectory deviates from this internalized baseline, users readily interpret the divergence as unreasonable or opportunistic behavior. 
Because the platform interface displays raw trajectories without revealing the underlying algorithmic dispatch logic, such as multi-order batching, pickup sequencing, platform prioritization, or sudden road obstacles, users are forced to infer worker intent from incomplete visual cues, transforming route anomalies into immediate accusations of ``taking a detour.'' 
For instance, U21 recalled tracking an order where an approaching courier abruptly reversed direction and remained stationary elsewhere, concluding with frustration that the worker was \textit{``still delivering my order, but he had inserted someone else's order in between, playing tricks.''}
However, what users read as strategic insubordination often masks necessary adaptation to real-world friction. 
As F5 explained, \textit{``the scene I encounter in reality and the scene shown on the user's phone are completely different,''} noting that when physical routes are blocked by congestion, choosing an alternative path registers on the customer interface merely as an inexplicable \textit{``bent line.''} 
This lack of contextual transparency encourages users to view standard logistical adjustments as rule-breaking maneuvers; as U17 asserted, couriers are \textit{``always trying to game the system''} by conveniently picking up unauthorized side orders. 
While F10 acknowledged that isolated violations do occur---leading to \textit{``especially winding''} trajectories driven by stacking too many orders---they emphasized that most irregular routes stem from unexpected operational constraints rather than \textit{``bad intentions.''}

\subsubsection{Staged Status Nodes as Interpretive Framework for Delivery Progress}

The staged status nodes of food delivery platforms provide users with a primary framework for understanding the delivery process, but they simultaneously constrain how that process is interpreted. 
By compressing a complex, exception-filled delivery process into a linear sequence, such as order accepted, meal picked up, delivering, and delivered, these milestones offer users a simplified mental model. 
As U12 observed, this progression functions \textit{``like a flowchart''} where orders should \textit{``move forward step by step,''} creating an implicit rule that if an order \textit{``gets stuck at one stage for too long, that order usually has a problem.''}
Similarly, U18 noted how these sequential markers impart a vivid \textit{``sense of order''} that allows users to \textit{``predict what is going to happen next''} and arrange their waiting routines accordingly. 
However, this rigid stage-based structure creates a default expectation of smooth, uninterrupted pacing. 
When a milestone repeats or stalls, users immediately perceive the deviation as abnormal, triggering anxiety and apportioning blame. 
Because the courier remains the most visible human actor in the transactional chain, they become the primary target of frustration when stage transitions falter. 
For instance, U17 recounted wondering if a courier was \textit{``slacking off''} or eating a meal simply because an order remained at the restaurant stage during a non-rush hour, while U21 described frustration when an unaccepted order left tracking static for half an hour. 
Consequently, even when structural bottlenecks originate from restaurant backlogs or platform dispatching delays, couriers inevitably absorb the fallout; as F13 noted, regardless of where an anomaly occurs in the chain, \textit{``I am the person that users can most easily reach to, so I am always the one being contacted.''}

\subsection{Managing the Data Shadow: How Couriers Internalize Interface Risk}

In parallel, couriers do not experience tracking interfaces as neutral informational displays; rather, they internalize user-facing platform metrics as a continuous risk environment. 
Anticipating how customers will interpret interface representations, couriers actively monitor and adjust their operational behaviors to mitigate complaints and manage their data shadows.

\subsubsection{Anticipatory Adjustments for Interface Legibility}

During delivery, couriers focus continuously beyond actual task progress, actively anticipating how their movements will translate into digital states on the user interface. 
When foreseeing interface events such as location stagnation or ETA extension, couriers proactively modify their routing or pacing to preempt customer misunderstandings. 
Because workers recognize that user judgments are mediated through platform traces rather than direct observation, managing the legibility of their digital footprint becomes a core component of everyday labor. 
As F1 explained, even during normal deliveries, couriers find themselves constantly calculating \textit{``how users will interpret my current state,''} ensuring that necessary pauses are \textit{``shorten[ed] as much as possible''} so that stationary moments do not invite suspicion on the map. 
Similarly, F2 noted that when restaurant delays occur, sending proactive messages helps alleviate customer anxiety \textit{``just because the status has not updated for a long time.''}

Consequently, maintaining interface legibility competes with task efficiency as an operational priority, compelling couriers to optimize the very data shadow they project over time.
Reflecting on this shift, F6 observed that while their early delivery work focused purely on speed, experience teaches the necessity of \textit{``looking normal''}---deliberately smoothing out trajectories so they appear \textit{``always moving forward''} rather than taking erratic pauses that read as problems to the users. 
This representational management frequently supersedes geographic optimization; as F9 noted, despite knowing faster local shortcuts, couriers often strictly follow default platform navigation because passing through signal-dead zones causes them to \textit{``disappear from the map and the user cannot see me,''} a risk of invisibility they refuse to take.

\subsubsection{Data Anomalies as Preemptive Complaint Triggers}

Through long fieldwork, couriers develop an acute awareness of the kinds of tracking situations that most often make customers anxious and lead to formal complaints.
These include long periods where an order looks stuck, lingering near a destination without showing completion, or repeated changes that extend the ETA. 
Because they understand where the system can fail, couriers stay more alert and act early when an order starts to reach a risky tracking pattern. 
Having repeatedly dealing with the consequences of user misunderstandings, couriers become better at noticing problems before they grow.
As F6 observed, being stationary for a long time and waiting near a drop-off location are \textit{``the two situations most likely to be complained about,''} so workers become \textit{``especially careful''} when an order gets close to those points---either by \textit{``mov[ing] a bit''} or contacting the customer \textit{``in advance.''} 
Similarly, F12 emphasized that while certain everyday fluctuations in the system still feel very \textit{``sensitive to users,''} such as gradually increasing delivery times or sudden location freezes.
Because of this, couriers must stay constantly \textit{``alert inside''} to prevent trouble from starting.

This expectation of metric-driven conflict directly alters how couriers prioritize tasks in real time, frequently overriding physical distance optimization in favor of data stabilization. 
For instance, F3 noted that even when standard routing strategy would suggest combining deliveries, a nearby time deadline may mean they must \textit{``finish this one first''} to avoid user panic. 
F13 explained that they weigh routing decisions against the risk of customer reactions. 
Sometimes they give priority to orders that \textit{``look more dangerous,''} even if they are farther away, because certain tracking states are more likely to catch a customer's attention and lead to urgent messages.

\subsection{Systemic Expectations: Bridging the Asymmetry Between Users and Couriers}
\label{subsec:Expectations}

Through long-term interaction with the platform, users and couriers not only develop understandings of the current system, but also gradually form stable expectations about ``how the system should operate.'' These expectations emerge from repeated misunderstandings, conflicts, and coordination practices in concrete experiences of use, and constitute a response to the inadequacies of the current structure of data presentation.

\subsubsection{User Demands for Contextual Transparency and Evidentiary Clarity}

Users’ expectations of future platform designs center primarily on mitigating the uncertainty during the waiting process.
Because current tracking interfaces supply decontextualized metrics---such as raw locations, rigid ETAs, and unannotated route lines---without explaining operational anomalies, users are left to independently speculate on the causes of delay. 
To alleviate this friction, participants consistently emphasized the need for system-verified contextual transparency. 
As U3 argued, anxiety would decrease if the platform \textit{``could directly tell me whether they were waiting for the food or stuck in traffic,''} noting that the core frustration stems from a system that \textit{``says nothing, and you can only think it through by yourself.''} 
Similarly, U11 clarified that occasional delays are entirely tolerable provided the platform provides operational visibility, explaining, \textit{``I do not mind it being a little late, but I want to know why it is late; if only the time changes, without any reason, I feel the process is not very reliable.''}
Beyond qualitative explanations, users expect platforms to establish transparent, standardized evidentiary metrics rather than forcing them to rely on subjective intuition. 
Pointing to the ambiguity of current timeout handling, U12 advocated for unambiguous system thresholds, such as a \textit{``clear record''} defining \textit{``how many minutes beyond the estimate counts as a punishable delay,''} to eliminate the need for personal guesswork. 
Reinforcing this demand for accountability infrastructure, U7 suggested that detailed temporal breakdowns, like tracking \textit{``when it started slowing down and how long it stayed slow,''} would furnish reliable evidence during disputes \textit{``instead of only being able to rely on my feelings.''}

\subsubsection{Courier Demands for Representational Fidelity and Relief from Explanatory Labor}

In contrast to users, couriers' systemic expectations focus heavily on mitigating the structural misunderstandings driven by divergences between platform data and actual delivery conditions. 
Because users routinely treat technical glitches or algorithmic latency as direct evidence of worker misconduct, couriers find themselves unfairly bear the emotional and communicative fallout of data mismatches.
Consequently, couriers urgently demand interface representations that faithfully mirror actual conditions to protect them from unmerited suspicion. 
Highlighting the friction of false digital representations, F6 pointed out that when normal transit registers as inactivity, \textit{``as soon as users see that, they will have complaints---this is actually a system problem, but in the end I am the one who bears it.''}
F4 echoed this operational vulnerability, describing how momentary GPS freezes make active movement look like an unauthorized stop, concluding that \textit{``if the record were a bit more accurate in this kind of situation, I would not always be suspected.''}
To reduce the constant burden for communication placed on couriers, they also expect platforms to automate the sharing of additional delivery information.
Instead of making workers repeatedly explain everyday problems---like delays from a busy kitchen or long waiting times---F2 argued that automated platform updates would remove the need to ``keep sending messages to explain.'' This, they said, would help prevent misunderstandings 
Concurring with this call for automated provenance tracking, F8 imagined an interface that automatically records operational milestones, such as \textit{``when I arrived at the restaurant and when I left,''} so that customers can inspect the unvarnished logistical timeline \textit{``and I would not have to keep explaining all the time.''}

\subsection{Situating Findings within the Chinese Delivery Context}
We note that certain findings appear closely tied to the local platform and labor environment. 
For example, as discussed above, some users relied heavily on estimated delivery times when organizing their expectations and attributing responsibility for delays. 
This tendency may be reinforced by locally documented expectations of rapid delivery and by platform mechanisms through which customer complaints can expose couriers to fault attribution and financial penalties~\cite{DengTangLai2026}. 
Similarly, couriers' anticipatory vigilance toward complaint risk is shaped by the severity of these punitive consequences, an influence that may not generalize to platforms with more lenient dispute-resolution mechanisms~\cite{Kusk2022}. 
Other findings appeared less directly tied to factors that participants or prior literature explicitly identified as specific to the Chinese context; however, this study did not collect comparative data outside China, so their transferability to other platform and cultural settings remains an open empirical question. For instance, users' tendency to read location data, route trajectories, and status nodes as behavioral evidence may arise partly from a broader interface condition---outcomes displayed without sufficient causal explanation---that is not unique to Chinese platforms, though we did not test this directly. 
Participants' expectation that the system provide contextual explanations and traceable records may reflect a response to information asymmetry more broadly, though we cannot rule out that this expectation is also shaped by local platform governance norms. 
It is worth noting that couriers' anticipatory adjustment of their trajectories so as to appear reasonable on the user-facing map is also a response to the structural condition of being represented through a user-facing interface, a condition that may also arise in other dual-sided tracking architectures. 
However, cultural factors specific to China may amplify this behavior among couriers.
We offer this distinction as an interpretive observation grounded in our own data and existing literature, not as a cross-national comparison.
This study did not collect data outside China, and the claims made above about generalizability are therefore necessarily provisional, intended only to open discussion of transferability; further evidence is left to future research.

\section{Discussion}

The delivery platform is more than a labor marketplace; it is a ubiquitous computing infrastructure that continuously transforms contingent physical activity into interface representations. Our findings reveal that tensions within food delivery are not reducible to interpersonal friction alone; they are also shaped and amplified by the design of real-time tracking systems and the patterns of visibility these systems produce.
In this section, we deconstruct the burden-shifting mechanism, the paradox of the data shadow, and the implications for designing socially responsible ubiquitous systems.

\subsection{Strucural Misalignment and the Paradox of Real-Time Visibility}

By comparing user and courier accounts, we demonstrate that delivery tensions extend beyond simple friction between impatient users and careless couriers. These tensions are also shaped and amplified by the design of real-time, location-based tracking systems, which translate situated movement, waiting, and prediction into visible cues of accountability.
Consequently, our contribution extends beyond the study of platform labor.

Prior work has shown that food delivery platforms rely on minimal, quantifiable traces such as GPS points to represent and manage labor, despite the work being materially and socially more complex than the map suggests~\cite{ShaikhEtAl2024}. 
Related research similarly shows that platform-mediated delivery relies heavily on human intervention, functioning more as a patchwork process than a seamless flow captured entirely by interface outputs~\cite{DalalEtAl2023}.
Building on this literature, we argue that real-time tracking systems do not simply reveal delivery progress. Instead, their design shapes what becomes visible, what remains invisible, and who is held accountable when digital representations fail to align with situated practice.

\subsubsection{Visibility Without Causation}

Our findings refine the transparency paradox by identifying a specific failure in ubiquitous systems: visibility without causation.
While classic Ubicomp literature treats visibility as a resource for coordination and social awareness~\cite{Dourish1992,Schmidt2002}, our study demonstrates that in real-time tracking, visibility becomes a liability when outcomes are detached from their underlying causes.
Building on Tang et al.'s~\cite{tang2010rethinking} frameworks of social- and purpose-driven location sharing, our findings reveal a specialized, defensive facet of the latter: accountability-driven sharing. 
While visibility in traditional purpose-driven sharing facilitates practical coordination, this accountability-driven facet emerges within asymmetrical, transactional environments where data traces are leveraged for behavioral evaluation.

This paradox stems from an information architecture that reduces the delivery process to a small set of visible outcomes. 
The four primary data modalities presented to users, i.e., static GPS markers, precise ETA countdowns, route visualizations, and linear status nodes, make the outcomes of the delivery process highly visible while obscuring the operational causes.
For instance, a static GPS dot accurately reflect a courier's lack of spatial movement, but the interface fails to communicate the real-world frictions producing that halt, such as restaurant delays, restricted building access, elevator queues, or signal loss. 
This reduction aligns with recent research showing that delivery platforms flatten complexity labor into simplified digital traces~\cite{ShaikhEtAl2024}, leaving ad hoc human coordination outside the formal platform representation~\cite{DalalEtAl2023}.

A similar issue occurs in the temporal domain. 
When ETAs are presented as rigid, minute-by-minute countdowns, users naturally interpret algorithmic prediction as firm commitment. 
Research on arrival-time systems suggests that while point forecasts can create misleading expectations of certainty, interval forecasts communicate uncertainty more effectively~\cite{Rollwagen2025}. 
Our findings reveal a related dynamic: users treat ETA drift not as algorithmic uncertainty, but as evidence that the courier failed to fulfill an obligation.
The interface provides enough precision to invite judgment, but insufficient context to support fair interpretation.

When interfaces provide granular data without contextual explanations, users fill the informational gap with behavioral attributions. 
Consistent with the correspondence bias, where people over-attribute outcomes to individual responsibility while underweighting situational constraints~\cite{GilbertMalone1995}, users frequently translated system-induced delays or environmental frictions into moralized judgments of courier negligence or opportunism.
The emphasis on visual transparency thus creates a sociotechnical blind spot: it compresses a complex labor process into a simplified narrative, turning ambiguous data anomalies into triggers for complaint and blame. 
This aligns with broader arguments that transparency does not automatically foster understanding and can backfire by creating mistrust or resistance~\cite{HuEtAl2024}.

The real-time nature of these updates aggravates this causal fracture. 
In non-real-time systems, users have the cognitive space to seek explanations. 
However, the continuous stream of mobile tracking data creates a truth-of-the-moment effect~\cite{sun2019your}. 
The interface's immediacy compresses the space for reflection, encouraging users to interpret every stall or deviation as a definitive behavioral signal rather than a stochastic environmental event.

Because surfacing the high-dimensional data behind every logistical delay is technically and cognitively infeasible, current interfaces default to false precision.
This resonates with Ubicomp discussions on uncertainty communication~\cite{kay2016ish}, demonstrating that point forecasts in dynamic mobile contexts often backfire.
Ultimately, visibility alone is insufficient.
Unless the system provides enough contextual intelligibility to ground visible traces in the realities of the work, increased visibility merely redistributes the interpretive burden to users and the blame burden to workers.

\subsubsection{Anticipatory Data Management and the Data Shadow}

We introduce \textit{anticipatory data management} as a theoretical extension of anticipatory compliance~\cite{Bucher2021}.
While Bucher et al.~\cite{Bucher2021} describe anticipatory compliance as workers aligning their overall work patterns to satisfy backend algorithms, anticipatory data management is a real-time tactical response to the user-facing representational layer.
It refers to workers' deliberate effort to align physical actions with expected interface outputs.
Couriers do not simply complete deliveries; they also manage their \emph{data shadow}, a visible, evaluative representation of their work that they only partially control, yet which dictates how they are judged.
This management includes adjusting routes to look reasonable on the customer's map, maintaining visible movement to avoid appearing idle, reordering deliveries based on which order looks more suspicious on the interface, and preemptively messaging users when the platform display invites negative interpretation.


This dynamic reveals a deep design paradox in ubiquitous sensing: while increased visibility is often framed as a means to improve coordination, the awareness of being monitored in real-time by an evaluative other causes workers to optimize for the appearance of the data rather than the efficiency of the task.Couriers are not merely trying to satisfy platform performance targets; they are actively avoiding being misread through traces that users interpret as evidence of motive and effort. 
While this shares surface similarities with related constructs, anticipatory data management is theoretically distinct along three dimensions. 
First, unlike \emph{impression management}, which concerns self-presentation through controllable verbal, facial, and behavioral cues in direct interpersonal encounters~\cite{chawla2021person}, anticipatory data management is mediated entirely by sensor-derived traces (GPS points, timestamps, route logs) that couriers can only partially and indirectly shape, and that are interpreted by an absent, asynchronous audience rather than a co-present observer. 
Second, unlike \emph{emotional labor}, which involves the regulation and display of emotion during service encounters~\cite{gabriel2023acceleration}, anticipatory data management targets the physical and spatial dimensions of work itself, such as route shape, pacing, and stopping behavior, rather than affective expression. 
Third, unlike \emph{anticipatory compliance}, which describes workers aligning their overall behavioral patterns with platform algorithms to satisfy backend performance metrics~\cite{Bucher2021}, anticipatory data management is a real-time, tactical response to how physical action will be rendered on the user-facing interface, independent of whether that rendering affects the courier's algorithmic score. 
Taken together, these distinctions position anticipatory data management as a response to a public, representational, and non-interpersonal audience, rather than to a co-present observer, an internal emotional display rule, or a backend scoring system.

The theoretical novelty of anticipatory data management lies in its focus on the representational layer of platform labor. 
This phenomenon is a manifestation of Goodhart’s Law in a ubiquitous context: when a measure (location/ETA) becomes a target (performance score), it ceases to be a good measure. 
Unlike office dashboards or smart health gamification where shared visibility foster motivation, this real-time feedback loop in gig labor leads to distorted physical actions. 
By prioritizing looking busy on the map over taking efficient shortcuts, couriers demonstrate that the representational layer of platform labor has become a moralized site of hidden work.
Ironically, a system oriented around improving transparency is associated with less efficient behavior because workers must optimize for how the delivery looks rather than the delivery itself.

\subsubsection{Platform Design as a Burden-Shifting Mechanism}

Our analysis shows that platforms do not merely transmit information; the resulting interface configurations distribute interpretive labor and accountability in systematic ways.
Adopting an Infrastructure Studies perspective~\cite{plantin2018infrastructure}, 
we conceptualize this selective visibility as a ``burden-shifting mechanism,'' a structural outcome of the infrastructure's  design rather than the personal intent of individual engineers. Where this paper describes what platform design 'withholds,' 'reinforces,' or 'externalizes,' this language characterizes the observed structure and consequences of the current interface as experienced by users and couriers, not a verified claim about designers' intent, which our user- and courier-facing interview method cannot establish. While these interface characteristics may often result from practical constraints, such as usability requirements that prioritize high-visibility metrics over causal depth, the limited representation of situational context in user-facing interfaces has significant socio-technical consequences.
Despite the logistical data such a ``God's-eye view''  would require, this context remains unrepresented in the interface. Consequently, users are positioned to judge based on highly visible traces, yet lack the context necessary to judge fairly. 
In contrast, couriers possess the situated context but are forced to bear the communicative and emotional burden of explaining systemic anomalies.
Observing these outcomes substantiates our argument that the current infrastructure creates an asymmetric accountability burden, regardless of the platform designers' initial motivations.

This is not purely a technical constraint: some withheld context may reflect genuine usability or complexity limits rather than deliberate design (see Section 5.2.1). 
However, two observations suggest this cannot fully explain the asymmetry. 
First, several low-cost interventions participants wanted, such as interval-based ETAs or coarse status annotations, would plausibly draw on data types the platform already uses for dispatch, suggesting their absence is not solely a matter of technical complexity. 
Second, courier trajectories and batching logic are, by necessity, available to the platform for dispatch purposes; why they remain unavailable to users is a question our interview data cannot resolve, and we flag this as a limitation rather than a claim we can substantiate. 
We therefore use burden-shifting to characterize the observable consequence of leaving potentially available contextual information unrepresented. 
Our data cannot determine whether these interface configurations arise primarily from usability constraints, information complexity, organizational priorities, commercial incentives, or deliberate design choices.

Structural frictions, such as restaurant backlog, access barriers, and signal instability, are only weakly represented on the interface. 
Instead, the burden of repairing misunderstanding is shifted onto couriers, who must message, call, justify, and sometimes alter their behavior to prevent complaints. 
When a delivery is delayed by faulty dispatching or kitchen bottlenecks, the infrastructure remains invisible and thus unaccountable;
the user perceives only the visible failure of the courier. 
This infrastructure-mediated accountability shifts the cognitive and emotional labor of explanation downward, forcing couriers to manually apologize for systemic errors that originate within the platform's own dispatching and logistics systems.
This automation by omission~\cite{VertesiEnriquez2025} illustrates how technical transparency can function as a mechanism of cost externalization, in which the core algorithm remains structurally outside the reach of user scrutiny while social conflict is displaced onto the human actors at the network's edge.

This mechanism also clarifies our RQ3 findings: users and couriers voiced different needs but pointed to the same structural asymmetry, users wanting system-verified reasons rather than speculation, couriers wanting the platform to assume explanatory work currently left to manual clarification, and both asking that explanatory responsibility be reallocated to the platform. Rather than treating visibility as inherently beneficial, we show it becomes a source of misalignment when detached from causal intelligibility; rather than viewing worker adaptation purely as algorithmic compliance, we show workers actively manage the representational layer of their labor; and rather than reading blame as simple impatience, we show how the interface arrangement is associated with who must interpret, explain, and be held accountable when representation diverges from practice.

\subsection{Design Directions for Accountable Real-Time Tracking Systems}
We emphasize that the design directions outlined below represent empirically derived design requirements---grounded directly in the interpretive misalignments identified in Section 4---and serve as untested conceptual proposals rather than evaluated technical artifacts.

Rather than presenting universally beneficial fixes, we frame our design implications as bounded directions with inherent feasibility constraints and side effects.
Because transparency is not automatically beneficial and interface redesign alone cannot resolve deeper inequities in platform labor~\cite{HuEtAl2024}, our goal is to identify how platforms can more responsibly support interpretation, coordination, and accountability.
These design directions are socio-technical design requirements, instead of speculative technical blueprints, grounded in the interpretive misalignments identified in our empirical results (see Section~\ref{subsec:Expectations}).
Their concrete feasibility, usability, and downstream effects (e.g., whether interval ETAs reduce or merely relocate user anxiety) remain open empirical questions.
They aim to redistribute explanatory labor from individual couriers back to the platform, addressing the specific informational needs voiced by participants.

\subsubsection{From Bare Visibility to Contextual Transparency}

The first design direction moves from bare visibility to contextual transparency using uncertainty-aware representations and event-triggered context.
Derived directly from the interpretive misalignments identified in our interviews (RQ3); this approach addresses users' desires for system-verified explanations (U3, U11, U12) and couriers' needs to reduce their explanatory burden (F2, F8).
Because current systems reveal outputs without their underlying conditions, a better design would expose selective, system-verifiable context without overwhelming users with operational detail.

Regarding ETAs, instead of rigid second-by-second countdowns frame predictions as commitments, platforms should implement interval forecasting.
By leveraging real-time variance in preparation times and collective traffic sensor data, the interface could display a bounded arrival window (e.g., ``15–22 minutes''). 
This communicates that an ETA is a dynamic estimate shaped by evolving conditions, not a fixed promise. Research confirms that interval forecasts improve uncertainty communication and are preferred when actual arrivals diverge from point estimates~\cite{Rollwagen2025}. 

Similarly, when routes deviation or prolonged stationary statuses occur, the interface should provide coarse-grained contextual annotations. 
By processing accelerometer and barometer data from the courier's smartphone, platforms can distinguish between road transit, vertical movement, and indoor waiting. 
For example, if a courier remains stationary for three minutes, the map dot could transform into a context-aware status icon (e.g., ``waiting at elevator''). 
This addresses the ambiguity of a static dot by providing verified situational context.
Several Chinese delivery platforms have started to count the time couriers spend waiting at traffic lights toward their allowed delivery time~\cite{Xinhua2026ChinaEconomicRoundtable}, however, this context has not incorporated into the user-end interfaces yet.

To make this concrete, the ETA display itself could be restructured as a two-segment bar rather than a single countdown: a \textit{preparation} segment derived from the restaurant's live order queue, and a \textit{transit} segment derived from real-time traffic and courier speed.
A brief tap on either segment would reveal its source (e.g., ``4 orders ahead of yours at the restaurant''), so that a delay is visibly attributable to a specific stage rather than treated as an undifferentiated broken promise. 
Likewise, the status icon described above need not be a single flag; a small, fixed vocabulary of icons (e.g., building access, elevator wait, signal gap, restaurant queue) mapped to sensor-inferred states would let users distinguish these situations at a glance, rather than receiving only a binary moving/stopped signal.
In the meantime, contextual transparency is not about maximal disclosure; users do not want endless detail, but a reliable basis for understanding anomalies. 
By allowing users to tap a status display to access a system-verified reason, the platform uses existing data to shoulder the explanatory burden.

However, this design direction has clear constraints. 
Contextual annotation relies on platform-side inference, which can be technically unreliable, and excessive operational detail may induce information overload or heighten anxiety.
As recent research warns, transparency becomes counterproductive when it exceeds meaningful interpretation~\cite{HuEtAl2024}. 
Therefore, contextual transparency must be selective and event-triggered, designed to explain high-risk anomalies rather than turning every delivery into a fully instrumented operational dashboard.

\subsubsection{Redistributing the Explanatory Burden to Platforms}

The second design direction redistributes the explanatory burden back to platforms. 
Currently, couriers operate as the default repair layer for interface ambiguity---an inefficient, unfair, and fragile dynamic. 
Platforms should automate this repair layer by recognizing system-induced anomalies.
When delays originate in restaurant preparation, dispatching logic, or building access, the platform is better positioned than the courier to provide a trustworthy explanation.
For instance, if a backend batching decision significantly delays an order, the system should proactively push a verified notification directly to the user. 
This transfers the burden of proof from the courier's manual messages to the platform's official interface.

Furthermore, post-hoc dispute processes should rely on system logs that reconstruct events from platform data, eliminating the need for unpaid, after-the-fact explanations from workers.
Strongly anchored in our RQ3 findings, this aligns with users' desires for clear evidentiary standards and couriers' requests for platform-led explanations.
However, to prevent automated explanations from becoming another opaque layer of control, these mechanisms must remain contestable.
If automated explanation becomes another opaque institutional layer that workers cannot question, it may simply replace one asymmetry with another. 
Worker-centered design research shows that workers highly value the ability to contextualize and challenge data through annotations~\cite{Stein2023}. 
Therefore, platform-generated explanations must support review and appeal.

To build a socially responsible infrastructure, we propose a \textit{report-difficulty} feature allowing couriers to flag real-time obstacles (e.g., road closures). 
The system could cross-reference this flag with data from nearby couriers or historical sensor traces to issue an official endorsement to the customer, granting the worker an algorithmic alibi. 
Concretely, the feature would offer preset categories (e.g., building access blocked, restaurant delay, road closure, no signal) matching situations our participants described (e.g., F5's blocked road, U11's signal-dead campus). 
A ``verified'' tag would only appear if another nearby courier's trace corroborates the same condition within a set time and radius.
This directly resolves the frustration courier expressed (e.g., F2, F8) regarding the unnecessary labor and misinterpretation risks associated with manual explanations.

This corroboration mechanism, however, is also the proposal most exposed to misuse. 
Its threshold could be gamed through peer coordination, a practice our findings and prior literature already document among couriers~\cite{Chen2022,Kusk2022}, while isolated couriers without nearby peers may fail to get corroboration for genuine delays. 
An unverified claim may also come to read as more suspicious than no claim at all. We therefore treat this proposal as the most speculative of the three, and return to its feasibility below.

\subsubsection{Data Mirroring and Dual-Sided Design}

The third design direction advocates designing for both sides of the accountability dyad. 
Current tracking systems are overwhelmingly user-facing, treating courier primarily as sources of movement data.
Yet, our findings reveal that couriers are deeply affected by these representations and attempt to blindly reconstruct them through experience.
One intervention is selective data mirroring via a lightweight overlay on the courier's navigation screen, providing a preview of the customer's view (e.g., current ETA, movement status). 
If a courier sees they are flagged as stalled due to GPS drift, they could proactively trigger a signal interruption status to manage the user's interpretation.

While mirroring supports symmetric awareness and reduces the effort of blindly managing representation, it involves real tensions.
Increased visibility could intensify self-monitoring for couriers, and platforms might appropriate mirroring for performance optimization rather than worker support.
Therefore, dual-sided design should not enforce total interface symmetry or constant exposure.
Instead, it should feature selective previews, anomaly alerts, limited mirroring of high-risk states, and worker control over visibility.
As worker-centered research demonstrates, expanded data access is only beneficial when workers retain the agency to contextualize, challenge, and govern how that data is used~\cite{Stein2023}.

\subsubsection{Institutional Feasibility, Incentives, and Design Scope}

It is important to acknowledge that platforms may lack motivation to implement these proposals, despite their technical feasibility using existing sensors and API data.
While uncertainty-aware ETAs or contextual tags could easily integrate into existing infrastructures, tools like contestable logs or data mirroring redistribute informational power and will likely face institutional resistance.
Organizational research suggests these asymmetries may be structurally advantageous to platforms, displacing the costs of coordination and repair onto the workers~\cite{VertesiEnriquez2025}.

Consequently, our design directions are conditional interventions, not straightforward fixes. 
They illustrate how interfaces could responsibly allocate interpretation, but they do not assume platforms will prioritize fairness over efficiency or managerial control. 
Design can mitigate misinterpretation, but it cannot resolve broader issues of labor precarity or asymmetric business models. 
Recognizing these institutional constraints, we frame our proposals as structural requirements for accountability rather than guaranteed implementations, highlighting exactly where current infrastructures fail to support social responsibility.
Ultimately, while these three directions map out a path toward contextual transparency and shared accountability, they remain conceptual proposals requiring future empirical and technical evaluation.

We distinguish two separate sources of risk across these proposals. 
Institutional reluctance, meaning platforms declining to redistribute informational power, applies most to contestable logs and data mirroring. 
Mechanism-level gameability, meaning the corroboration threshold itself being exploitable, as discussed above, applies specifically to the algorithmic alibi. These are not the same objection: a feature could clear institutional adoption and still fail on gameability, or vice versa. 
Treating them separately clarifies that redesigning the corroboration mechanism (e.g., weighting reports by a courier's own historical reliability, or requiring corroboration from couriers outside the same delivery batch to reduce collusion incentives) is a distinct, and possibly more tractable, problem than the broader question of whether platforms are willing to cede informational control at all.


\subsection{Limitations, Transferability, and Future Work}

Despite its contributions, this study has several limitations.
First, it relies on retrospective semi-structured interviews. 
While this method captures participants' interpretations rather than real-time observation, it is a widely adopted method for investigating the interpretive meaning-making processes that drive accountability judgments~\cite{Shaikh2023,Kusk2022}.
System logs can objectively document what happened, but they are structurally incapable of surfacing why---the moralized judgments, anticipated consequences, and cognitive loads shaping these misalignments. 
Qualitative interviews are therefore essential for accessing these unobservable dimensions.
Because our findings rely on retrospective interviews, participants may have reconstructed past events in light of later experiences. 
This concern is particularly relevant to couriers’ accounts of workarounds, multi-order delivery, route adjustments, and other practices that could be perceived as inefficient or rule-bending, which may have been selectively withheld, softened, or retrospectively rationalized. 
We therefore treat these accounts as situated interpretations rather than complete behavioral records and encourage future research combining interviews with direct observation, platform logs, or matched user–courier cases.
Second, while our methodology sampled perspectives from both sides of the delivery ecosystem, our dataset does not contain matched user-courier pairs evaluating the exact same delivery incident in real time. 
While retrospective dual-perspective interviews successfully exposed systemic misalignments, future work should deploy in-situ event-tracing protocols to capture synchronous, matched dyadic interpretations of specific tracking anomalies.
Third, our findings are grounded in the Chinese food delivery context; norms of punctuality, labor culture, and complaint behaviors naturally vary across settings. 
Finally, our analysis focused primarily on the user-courier dyad. 
While this isolated a specific interface-level accountability problem, it excludes other critical actors in the ecosystem, such as restaurant staff, dispatch teams, and platform designers.


Transferability, however, is not an all-or-nothing question.
The tendency for simplified traces to invite accountability judgments under uncertainty likely reflects the structural features of real-time tracking systems globally~\cite{lu2026ubiquitous}. 
Research in other contexts confirms that platforms rely on minimal location traces to represent labor~\cite{ShaikhEtAl2024} and depend on situated human interventions that formal interfaces fail to capture~\cite{DalalEtAl2023}.
These parallels suggest our identified mechanisms may travel across regions, even if their specific manifestations vary.

Future comparative research should test which findings generalize across countries and platforms and which remain applicable to local labor cultures.
A broader stakeholder perspective is needed to examine how interpretive misalignments are institutionally produced and sustained.
Methodologically, we will also combine qualitative inquiry with in-situ techniques, such as diary studies or interface-event logging, to clarify precisely when data states trigger judgment and behavioral alteration.
We also plan to incorporate the perspectives of platform engineers to better understand the design intent behind the burden-shifting mechanism.

Overall, our findings suggest that real-time tracking in food delivery is not a neutral coordination layer; it is an accountability infrastructure. By leaving causes largely unrepresented while making outcomes visible, this infrastructure is associated with users judging through decontextualized traces and couriers having to manage their data shadows. 
The next generation of ubiquitous systems must evolve into Infrastructures of Care, where systems take responsibility for the social consequences of their technical orchestration. 
The goal of accountable tracking is not maximal transparency, but an equitable distribution of interpretative labor.
To ground this in the Ubicomp tradition, future research must pursue technical prototype verification, utilizing sensor-fusion (e.g., barometer data for vertical movement) to technically resolve the interpretive misalignments diagnosed here.

\section{Conclusion}
Based on interviews with 23 users and 17 couriers on Chinese platforms, we find that tensions in food delivery are shaped by real-time tracking interfaces. 
Users operate within a data-as-behavior framework, translating interface traces into moralized judgments, while couriers engage in anticipatory data management to navigate the risk posed by their own data shadow. 
Together these point to a burden-shifting mechanism, in which current tracking architectures effectively shift onto couriers explanatory labor that could instead be supported by platform-held logistical data (though this may not be the intended purpose of the design). 
Participants' expectations converge on a shared demand: contextual transparency, so that platforms take on more of the explanatory burden currently borne jointly by users and couriers. 
We call for Infrastructures of Care that prioritize the equitable distribution of interpretive labor, ensuring data serves coordination rather than judgment.

\section*{Acknowledgments}
This work was supported in part by the Research Grants Council of Hong Kong under Grant No. 14201425.

\section*{Ethics Statement and AI Usage}

This work does not raise any ethical issues. 
Google Gemini 3.1 Pro was used to improve the quality of writing. 
The used prompt is: \textit{Please help me polish the following text:} [text].

\bibliographystyle{ACM-Reference-Format}
\bibliography{references}

@article{gabriel2023acceleration,
  title={The acceleration of emotional labor research: Navigating the past and steering toward the future},
  author={Gabriel, Allison S and Diefendorff, James M and Grandey, Alicia A},
  journal={Personnel Psychology},
  volume={76},
  number={2},
  pages={511--545},
  year={2023},
  publisher={Wiley Online Library}
}

@misc{Xinhua2026ChinaEconomicRoundtable,
  author = {{Xinhua}},
  title = {China Economic Roundtable: From algorithms to social protection -- How China is responding to challenges facing 10 mln delivery riders},
  year = {2026},
  howpublished = {\url{https://www.bignewsnetwork.com/news/279308966/china-economic-roundtable-china-focus-from-algorithms-to-social-protection-how-china-is-responding-to-challenges-facing-10-mln-delivery-riders}},
  note = {Accessed: 2026-09-29}
}

@article{chawla2021person,
  title={A person-centered view of impression management, inauthenticity, and employee behavior},
  author={Chawla, Nitya and Gabriel, Allison S and Rosen, Christopher C and Evans, Jonathan B and Koopman, Joel and Hochwarter, Wayne A and Palmer, Joshua C and Jordan, Samantha L},
  journal={Personnel Psychology},
  volume={74},
  number={4},
  pages={657--691},
  year={2021},
  publisher={Wiley Online Library}
}

@misc{meituan_q2_2025,
  title = {Meituan (HKEX: 3690) Q2 2025 Earnings Call},
  howpublished = {\url{https://www.alphaspread.com/security/hkex/3690/investor-relations/earnings-call/q2-2025}},
  year = {2025},
  note = {Accessed: 2026-07-13},
  publisher = {AlphaSpread}
}

@article{humphreys2007mobile,
  title={Mobile social networks and social practice: A case study of Dodgeball},
  author={Humphreys, Lee},
  journal={Journal of Computer-Mediated Communication},
  volume={13},
  number={1},
  pages={341--360},
  year={2007},
  publisher={Oxford University Press Oxford, UK}
}

@inproceedings{barkhuus2008awareness,
  title={From awareness to repartee: sharing location within social groups},
  author={Barkhuus, Louise and Brown, Barry and Bell, Marek and Sherwood, Scott and Hall, Malcolm and Chalmers, Matthew},
  booktitle={proceedings of the SIGCHI conference on human factors in computing systems},
  pages={497--506},
  year={2008}
}

@inproceedings{toch2010empirical,
  title={Empirical models of privacy in location sharing},
  author={Toch, Eran and Cranshaw, Justin and Drielsma, Paul H and Tsai, Janice Y and Kelley, Patrick G and Springfield, James and Sadeh, Norman},
  booktitle={Proceedings of the 12th ACM international conference on Ubiquitous computing},
  pages={129--138},
  year={2010}
}

@inproceedings{wiese2011close,
  title={Are you close with me? Are you nearby? Investigating social groups, closeness, and willingness to share},
  author={Wiese, Jason and Kelley, Patrick G and Cranor, Lorrie Faith and Dabbish, Laura and Hong, Jason I and Zimmerman, John},
  booktitle={Proceedings of the 13th international conference on Ubiquitous computing},
  pages={197--206},
  year={2011}
}

@inproceedings{wilson2013privacy,
  title={Privacy manipulation and acclimation in a location sharing application},
  author={Wilson, Shana and Cranshaw, Justin and Sadeh, Norman and Acquisti, Alessandro and Cranor, Lorrie Faith and Springfield, James and Balasubramanian, Anand},
  booktitle={Proceedings of the 2013 ACM international joint conference on Pervasive and ubiquitous computing},
  pages={549--558},
  year={2013}
}

@inproceedings{brown2007locating,
  title={Locating Family Values: A Field Trial of the Whereabouts Clock},
  author={Brown, Barry and Taylor, Alex S and Izadi, Shahram and Sellen, Abigail and Kaye, Jofish and Eardley, Richard},
  booktitle={International Conference on Ubiquitous Computing},
  pages={354--371},
  year={2007},
  organization={Springer}
}

@article{sadeh2009understanding,
  title={Understanding and capturing people’s privacy policies in a mobile social networking application},
  author={Sadeh, Norman and Hong, Jason and Cranor, Lorrie and Fette, Ian and Kelley, Patrick and Prabaker, Mads and Rao, Joel},
  journal={Personal and ubiquitous computing},
  volume={13},
  number={6},
  pages={401--412},
  year={2009},
  publisher={Springer}
}

@inproceedings{tang2010rethinking,
  title={Rethinking location sharing: exploring the implications of social-driven vs. purpose-driven location sharing},
  author={Tang, Kaimin P and Lin, Jenifer and Hong, Jason I and Siewiorek, Daniel P and Sadeh, Norman},
  booktitle={Proceedings of the 12th ACM international conference on Ubiquitous computing},
  pages={85--94},
  year={2010}
}

@article{huang2023algorithmic,
  author  = {Huang, Hua},
  title   = {Algorithmic management in food-delivery platform economy in China},
  journal = {New Technology, Work and Employment},
  year    = {2023},
  volume  = {38},
  pages   = {185--205},
  doi     = {10.1111/ntwe.12228}
}

@inproceedings{star1994steps,
  title={Steps towards an ecology of infrastructure: complex problems in design and access for large-scale collaborative systems},
  author={Star, Susan Leigh and Ruhleder, Karen},
  booktitle={Proceedings of the 1994 ACM conference on Computer supported cooperative work},
  pages={253--264},
  year={1994}
}

@inproceedings{kay2016ish,
  title={When (ish) is my bus? user-centered visualizations of uncertainty in everyday, mobile predictive systems},
  author={Kay, Matthew and Kola, Tara and Hullman, Jessica R and Munson, Sean A},
  booktitle={Proceedings of the 2016 chi conference on human factors in computing systems},
  pages={5092--5103},
  year={2016}
}

@article{preyss2007stochastic,
  title={Stochastic modeling of human learning behavior},
  author={Preyss, Albert E and Meiry, Jacob L},
  journal={IEEE Transactions on Man-Machine Systems},
  volume={9},
  number={2},
  pages={36--46},
  year={2007},
  publisher={IEEE}
}

@inproceedings{chalmers2004seamful,
  title={Seamful interweaving: heterogeneity in the theory and design of interactive systems},
  author={Chalmers, Matthew and Galani, Areti},
  booktitle={Proceedings of the 5th conference on Designing interactive systems: processes, practices, methods, and techniques},
  pages={243--252},
  year={2004}
}

@article{sun2019your,
  title={Your order, their labor: An exploration of algorithms and laboring on food delivery platforms in China},
  author={Sun, Ping},
  journal={Chinese journal of communication},
  volume={12},
  number={3},
  pages={308--323},
  year={2019},
  publisher={Taylor \& Francis}
}

@article{he2021demand,
  title={On-demand service delivery under asymmetric information: Priority pricing, market selection, and horizontal substitution},
  author={He, Qiaochu and Fan, Xiaoshuai and Chen, Ying-Ju and Yang, Hai},
  journal={International Journal of Production Economics},
  volume={237},
  pages={108146},
  year={2021},
  publisher={Elsevier}
}

@article{russo2023urban,
  title={Urban courier delivery in a smart city: the user learning process of travel costs enhanced by emerging technologies},
  author={Russo, Francesco and Comi, Antonio},
  journal={Sustainability},
  volume={15},
  number={23},
  pages={16253},
  year={2023},
  publisher={Mdpi}
}

@article{wen2023enough,
  title={Enough waiting for the couriers: Learning to estimate package pick-up arrival time from couriers’ spatial-temporal behaviors},
  author={Wen, Haomin and Lin, Youfang and Wu, Fan and Wan, Huaiyu and Sun, Zhongxiang and Cai, Tianyue and Liu, Hongyu and Guo, Shengnan and Zheng, Jianbin and Song, Chao and others},
  journal={ACM Transactions on Intelligent Systems and Technology},
  volume={14},
  number={3},
  pages={1--22},
  year={2023},
  publisher={ACM New York, NY}
}

@article{kang2025you,
  title={You're the One Whom I'm Talking To: The Role of Contextual External Human-Machine Interfaces in Multi-Road User Conflict Scenarios},
  author={Kang, Yumin and Park, Jeongju and Hwang, Seokhyun and Seong, Minwoo and Kim, Gwangbin and Kim, SeungJun},
  journal={Proceedings of the ACM on Interactive, Mobile, Wearable and Ubiquitous Technologies},
  volume={9},
  number={3},
  pages={1--37},
  year={2025},
  publisher={ACM New York, NY, USA}
}

@article{baseman2025clinical,
  title={Clinical Standards and Proximate Futures: Participatory Design Futuring of Diabetes Technologies with an Under-Resourced Community},
  author={Baseman, Cynthia M and Dembure, Seka M and Swinger, Nathaniel and Bondu, Muni T and DiSalvo, Betsy and Arriaga, Rosa I},
  journal={Proceedings of the ACM on Interactive, Mobile, Wearable and Ubiquitous Technologies},
  volume={9},
  number={4},
  pages={1--30},
  year={2025},
  publisher={ACM New York, NY, USA}
}

@article{liu2018foodnet,
  title={FooDNet: Toward an optimized food delivery network based on spatial crowdsourcing},
  author={Liu, Yan and Guo, Bin and Chen, Chao and Du, He and Yu, Zhiwen and Zhang, Daqing and Ma, Huadong},
  journal={IEEE Transactions on Mobile Computing},
  volume={18},
  number={6},
  pages={1288--1301},
  year={2018},
  publisher={IEEE}
}

@article{jarrahi2021algorithmic,
  title={Algorithmic management in a work context},
  author={Jarrahi, Mohammad Hossein and Newlands, Gemma and Lee, Min Kyung and Wolf, Christine T and Kinder, Eliscia and Sutherland, Will},
  journal={Big data \& society},
  volume={8},
  number={2},
  pages={20539517211020332},
  year={2021},
  publisher={SAGE Publications Sage UK: London, England}
}

@article{griesbach2019algorithmic,
  author  = {Griesbach, Kathleen and Reich, Adam and Elliott-Negri, Luke and Milkman, Ruth},
  title   = {Algorithmic Control in Platform Food Delivery Work},
  journal = {Socius: Sociological Research for a Dynamic World},
  year    = {2019},
  volume  = {5},
  doi     = {10.1177/2378023119870041}
}

@article{tuomi2024strategies,
  author   = {Tuomi, Aarni and Jianu, Bogdan and Hua, Min and Roelofsen, Maartje and Ascen{\c{c}}{\~a}o, M. Paula},
  title    = {Strategies for communicating and mitigating algorithmic control on delivery platforms},
  journal  = {Convergence: The International Journal of Research into New Media Technologies},
  year     = {2024},
  volume   = {30},
  number   = {5},
  pages    = {1685--1709},
  doi      = {10.1177/13548565241242685}
}

@article{duggan2023algorithmic,
  author   = {Duggan, James and Carbery, Ronan and McDonnell, Anthony and Sherman, Ultan},
  title    = {Algorithmic {HRM} control in the gig economy: The app-worker perspective},
  journal  = {Human Resource Management},
  year     = {2023},
  volume   = {62},
  number   = {6},
  pages    = {883--899},
  doi      = {10.1002/hrm.22168}
}

@article{Kusk2022,
author = {Kusk, Kalle and Bossen, Claus},
title = {Working with Wolt: An Ethnographic Study of Lenient Algorithmic Management on a Food Delivery Platform},
year = {2022},
issue_date = {January 2022},
publisher = {Association for Computing Machinery},
address = {New York, NY, USA},
volume = {6},
number = {GROUP},
url = {https://doi.org/10.1145/3492823},
doi = {10.1145/3492823},
journal = {Proc. ACM Hum.-Comput. Interact.},
month = jan,
articleno = {4},
numpages = {22}
}

@article{Shaikh2023,
author = {Shaikh, Riyaj and Lampinen, Airi and Brown, Barry},
title = {The Work to Make Piecework Work: An Ethnographic Study of Food Delivery Work in India During the COVID-19 Pandemic},
year = {2023},
issue_date = {October 2023},
publisher = {Association for Computing Machinery},
address = {New York, NY, USA},
volume = {7},
number = {CSCW2},
url = {https://doi.org/10.1145/3610034},
doi = {10.1145/3610034},
journal = {Proc. ACM Hum.-Comput. Interact.},
month = oct,
articleno = {243},
numpages = {23}
}

@article{Chen2022,
author = {Chen, Zhilong and Lan, Xiaochong and Piao, Jinghua and Zhang, Yunke and Li, Yong},
title = {A Mixed-Methods Analysis of the Algorithm-Mediated Labor of Online Food Deliverers in China},
year = {2022},
issue_date = {November 2022},
publisher = {Association for Computing Machinery},
address = {New York, NY, USA},
volume = {6},
number = {CSCW2},
url = {https://doi.org/10.1145/3555585},
doi = {10.1145/3555585},
journal = {Proc. ACM Hum.-Comput. Interact.},
month = nov,
articleno = {484},
numpages = {24}
}

@article{Yao2021,
author = {Yao, Zheng and Weden, Silas and Emerlyn, Lea and Zhu, Haiyi and Kraut, Robert E.},
title = {Together But Alone: Atomization and Peer Support among Gig Workers},
year = {2021},
issue_date = {October 2021},
publisher = {Association for Computing Machinery},
address = {New York, NY, USA},
volume = {5},
number = {CSCW2},
url = {https://doi.org/10.1145/3479535},
doi = {10.1145/3479535},
journal = {Proc. ACM Hum.-Comput. Interact.},
month = oct,
articleno = {391},
numpages = {29}
}

@inproceedings{Hernandez2024,
author = {Hernandez, Rie Helene (Lindy) and Song, Qiurong and Kou, Yubo and Gui, Xinning},
title = {"At the end of the day, I am accountable": Gig Workers' Self-Tracking for Multi-Dimensional Accountability Management},
year = {2024},
isbn = {9798400703300},
publisher = {Association for Computing Machinery},
address = {New York, NY, USA},
url = {https://doi.org/10.1145/3613904.3642151},
doi = {10.1145/3613904.3642151},
booktitle = {Proceedings of the 2024 CHI Conference on Human Factors in Computing Systems},
articleno = {382},
numpages = {20},
location = {Honolulu, HI, USA},
series = {CHI '24}
}

@article{Rosenblat2017,
  author  = {Rosenblat, Alex and Levy, Karen and Barocas, Solon and Hwang, Tim},
  title   = {Discriminating Tastes: Uber's Customer Ratings as Vehicles for Workplace Discrimination},
  journal = {Policy \& Internet},
  year    = {2017},
  doi     = {10.1002/poi3.153}
}

@inproceedings{Dourish1992,
author = {Dourish, Paul and Bellotti, Victoria},
title = {Awareness and coordination in shared workspaces},
year = {1992},
isbn = {0897915429},
publisher = {Association for Computing Machinery},
address = {New York, NY, USA},
url = {https://doi.org/10.1145/143457.143468},
doi = {10.1145/143457.143468},
booktitle = {Proceedings of the 1992 ACM Conference on Computer-Supported Cooperative Work},
pages = {107–114},
numpages = {8},
location = {Toronto, Ontario, Canada},
series = {CSCW '92}
}

@article{Schmidt2002,
  author  = {Schmidt, Kjeld},
  title   = {The Problem with `Awareness': Introductory Remarks on `Awareness in CSCW'},
  journal = {Computer Supported Cooperative Work (CSCW)},
  year    = {2002},
  volume  = {11},
  pages   = {285--298},
  doi     = {10.1023/A:1021272909573}
}

@article{RosenblatStark2016,
  author  = {Rosenblat, Alex and Stark, Luke},
  title   = {Algorithmic Labor and Information Asymmetries: A Case Study of Uber's Drivers},
  journal = {International Journal of Communication},
  volume  = {10},
  year    = {2016},
  pages   = {3758--3784}
}

@article{Cheon2025,
author = {Cheon, EunJeong and Erickson, Ingrid},
title = {Fulfillment of the Work Games: Warehouse Workers' Experiences with Algorithmic Management},
year = {2025},
issue_date = {November 2025},
publisher = {Association for Computing Machinery},
address = {New York, NY, USA},
volume = {9},
number = {7},
url = {https://doi.org/10.1145/3757409},
doi = {10.1145/3757409},
journal = {Proc. ACM Hum.-Comput. Interact.},
month = oct,
articleno = {CSCW228},
numpages = {30}
}

@misc{goodhart2015goodhart,
  title={Goodhart’s law},
  author={Goodhart, Charles},
  journal={The encyclopedia of central banking},
  volume={227},
  year={2015},
  publisher={Edward Elgar Publishing Cheltenham, UK}
}

@article{GilbertMalone1995,
  author  = {Gilbert, Daniel T. and Malone, Patrick S.},
  title   = {The Correspondence Bias},
  journal = {Psychological Bulletin},
  volume  = {117},
  number  = {1},
  pages   = {21--38},
  year    = {1995},
  doi     = {10.1037/0033-2909.117.1.21},
  pmid    = {7870861}
}

@article{JonesHarris1967,
  author  = {Jones, Edward E. and Harris, Victor A.},
  title   = {The Attribution of Attitudes},
  journal = {Journal of Experimental Social Psychology},
  volume  = {3},
  number  = {1},
  pages   = {1--24},
  year    = {1967},
  doi     = {10.1016/0022-1031(67)90034-0}
}

@article{HanLiuLoewenstein2023,
  author  = {Han, Yi and Liu, Yiming and Loewenstein, George},
  title   = {Confusing Context with Character: Correspondence Bias in Economic Interactions},
  journal = {Management Science},
  volume  = {69},
  number  = {2},
  pages   = {1070--1091},
  year    = {2023},
  doi     = {10.1287/mnsc.2022.4384}
}

@article{Ravula2023DeliveryPerformance,
  author       = {Prashanth Ravula},
  title        = {Impact of delivery performance on online review ratings: the role of temporal distance of ratings},
  journal      = {Journal of Marketing Analytics},
  year         = {2023},
  volume       = {11},
  number       = {2},
  pages        = {149--159},
  doi          = {10.1057/s41270-022-00168-5},
  pmcid        = {PMC9115748},
  publisher    = {Palgrave Macmillan},
}

@article{Quarles2025,
  author       = {Hillary Quarles and Gregory L. Simon},
  title        = {Platforming space: How food delivery platforms optimize users through physical, digital, and virtual space},
  journal      = {Digital Geography and Society},
  year         = {2025},
  volume       = {9},
  pages        = {100134},
  issn         = {2666-3783},
  doi          = {10.1016/j.diggeo.2025.100134},
  url          = {https://www.sciencedirect.com/science/article/pii/S2666378325000236},
}

@article{weinshel2025would,
  title={" I would still use it but I wouldn't trust it": Evaluating Mechanisms for Transparency and Control for Smart-Home Sensors},
  author={Weinshel, Ben and Agarwal, Yuvraj and Bauer, Lujo},
  journal={Proceedings of the ACM on Interactive, Mobile, Wearable and Ubiquitous Technologies},
  volume={9},
  number={2},
  pages={1--33},
  year={2025},
  publisher={ACM New York, NY, USA}
}

@inproceedings{ShaikhEtAl2024,
  author    = {Shaikh, Riyaj Isamiya and Singh, Anubha and Brown, Barry and Lampinen, Airi},
  title     = {Not Just a Dot on the Map: Food Delivery Workers as Infrastructure},
  booktitle = {Proceedings of the 2024 CHI Conference on Human Factors in Computing Systems},
  series    = {CHI '24},
  year      = {2024},
  publisher = {Association for Computing Machinery},
  address   = {New York, NY, USA},
  articleno = {385},
  pages     = {1--15},
  doi       = {10.1145/3613904.3641918}
}

@article{Zong2024,
  author  = {Zong, Yuanyuan and Tsaur, Sheng-Hshiung and Dai, You-Yu},
  title   = {Hassles of Platform-Based Food Couriers: An Asian Case Study},
  journal = {Journal of Transport \& Health},
  volume  = {34},
  pages   = {101743},
  year    = {2024},
  issn    = {2214-1405},
  doi     = {10.1016/j.jth.2023.101743}
}

@article{DengTangLai2026,
  author  = {Deng, T. and Tang, C. and Lai, Y.},
  title   = {How Workers Prevent Negative Online Reviews under Algorithmic Management: Evidence from Chinese Food-Delivery Platform},
  journal = {Information Technology \& People},
  volume  = {39},
  number  = {2},
  pages   = {1091--1115},
  year    = {2026},
  doi     = {10.1108/ITP-12-2023-1201}
}

@article{Xu2025,
author = {Xu, Haosu and Luo, Yiming and Wang, Yihong and Wang, Ding and Pan, Yushan and Cai, Shaoyu},
title = {From Routes to Ratings: Challenges and Strategies in Food Delivery Work: From Routes to Ratings: Challenges and Strategies in Food Delivery Work},
year = {2025},
issue_date = {Sep 2025},
publisher = {Kluwer Academic Publishers},
address = {USA},
volume = {34},
number = {3},
issn = {0925-9724},
url = {https://doi.org/10.1007/s10606-025-09525-1},
doi = {10.1007/s10606-025-09525-1},
journal = {Comput. Supported Coop. Work},
month = jul,
pages = {911–948},
numpages = {38}
}

@inproceedings{Hsieh2023,
author = {Hsieh, Jane and Karger, Miranda and Zagal, Lucas and Zhu, Haiyi},
title = {Co-Designing Alternatives for the Future of Gig Worker Well-Being: Navigating Multi-Stakeholder Incentives and Preferences},
year = {2023},
isbn = {9781450398930},
publisher = {Association for Computing Machinery},
address = {New York, NY, USA},
url = {https://doi.org/10.1145/3563657.3595982},
doi = {10.1145/3563657.3595982},
booktitle = {Proceedings of the 2023 ACM Designing Interactive Systems Conference},
pages = {664–687},
numpages = {24},
location = {Pittsburgh, PA, USA},
series = {DIS '23}
}

@inproceedings{Sekharan2025,
author = {Sekharan, Abhishek and Bui, Matthew and Hui, Julie},
title = {Designing for Instant Convenience: Dark Stores, Spatial Control and Worker Quantification in On-demand Delivery Platforms},
year = {2025},
isbn = {9798400714856},
publisher = {Association for Computing Machinery},
address = {New York, NY, USA},
url = {https://doi.org/10.1145/3715336.3735733},
doi = {10.1145/3715336.3735733},
booktitle = {Proceedings of the 2025 ACM Designing Interactive Systems Conference},
pages = {899–914},
numpages = {16},
location = {
},
series = {DIS '25}
}

@inproceedings{Lee2015,
author = {Lee, Min Kyung and Kusbit, Daniel and Metsky, Evan and Dabbish, Laura},
title = {Working with Machines: The Impact of Algorithmic and Data-Driven Management on Human Workers},
year = {2015},
isbn = {9781450331456},
publisher = {Association for Computing Machinery},
address = {New York, NY, USA},
url = {https://doi.org/10.1145/2702123.2702548},
doi = {10.1145/2702123.2702548},
booktitle = {Proceedings of the 33rd Annual ACM Conference on Human Factors in Computing Systems},
pages = {1603–1612},
numpages = {10},
location = {Seoul, Republic of Korea},
series = {CHI '15}
}

@article{DongZhangWu2025Burnout,
  author       = {Jian Dong and Guoyong Zhang and Lizhi Wu},
  title        = {Life against algorithmic management: a study on burnout and its influencing factors among food delivery riders},
  journal      = {Frontiers in Public Health},
  year         = {2025},
  volume       = {13},
  pages        = {1531541},
  doi          = {10.3389/fpubh.2025.1531541},
  pmid         = {40302769},
  pmcid        = {PMC12037544},
}

@article{Bucher2021,
  author  = {Eliane L{\'e}ontine Bucher and Peter Kalum Schou and Matthias Waldkirch},
  title   = {Pacifying the algorithm: Anticipatory compliance in the face of algorithmic management in the gig economy},
  journal = {Organization},
  volume  = {28},
  number  = {1},
  pages   = {44--67},
  year    = {2021},
  doi     = {10.1177/1350508420961531}
}

@article{plantin2018infrastructure,
  title={Infrastructure studies meet platform studies in the age of Google and Facebook},
  author={Plantin, Jean-Christophe and Lagoze, Carl and Edwards, Paul N and Sandvig, Christian},
  journal={New media \& society},
  volume={20},
  number={1},
  pages={293--310},
  year={2018},
  publisher={Sage Publications Sage UK: London, England}
}

@article{lu2026ubiquitous,
  title={Ubiquitous Lingering Technologies: What's Left Behind by the Past “Proximate Futures”?},
  author={Lu, Alex Jiahong and Sun, Yuling},
  journal={Proceedings of the ACM on Interactive, Mobile, Wearable and Ubiquitous Technologies},
  volume={10},
  number={1},
  pages={1--23},
  year={2026},
  publisher={ACM New York, NY, USA}
}

@article{zhang2019route,
  title={Route prediction for instant delivery},
  author={Zhang, Yan and Liu, Yunhuai and Li, Genjian and Ding, Yi and Chen, Ning and Zhang, Hao and He, Tian and Zhang, Desheng},
  journal={Proceedings of the ACM on Interactive, Mobile, Wearable and Ubiquitous Technologies},
  volume={3},
  number={3},
  pages={1--25},
  year={2019},
  publisher={ACM New York, NY, USA}
}

@article{mirjafari2019differentiating,
  title={Differentiating higher and lower job performers in the workplace using mobile sensing},
  author={Mirjafari, Shayan and Masaba, Kizito and Grover, Ted and Wang, Weichen and Audia, Pino and Campbell, Andrew T and Chawla, Nitesh V and Swain, Vedant Das and Choudhury, Munmun De and Dey, Anind K and others},
  journal={Proceedings of the ACM on Interactive, Mobile, Wearable and Ubiquitous Technologies},
  volume={3},
  number={2},
  pages={1--24},
  year={2019},
  publisher={ACM New York, NY, USA}
}

@article{kwon2018connected,
  title={The connected shower: Studying intimate data in everyday life},
  author={Kwon, Hyosun and Fischer, Joel E and Flintham, Martin and Colley, James},
  journal={Proceedings of the ACM on Interactive, Mobile, Wearable and Ubiquitous Technologies},
  volume={2},
  number={4},
  pages={1--22},
  year={2018},
  publisher={ACM New York, NY, USA}
}

@inproceedings{Rollwagen2025,
  author    = {Rollwagen, Alice and Horn, Alexander and Schmidtner, Stefanie and Riener, Andreas},
  title     = {Communicating Uncertainty in Arrival Time Predictions for Public Transport: A Comparison of Point and Interval Forecasts},
  booktitle = {Proceedings of Mensch und Computer 2025},
  year      = {2025},
  pages     = {515--519},
  doi       = {10.1145/3743049.3748542}
}

@article{xie2024,
  author  = {Si Xie and Siddhartha Sharma and Amit Mehra and Arslan Aziz},
  title   = {Strategic Expectation Setting of Delivery Time on Marketplaces},
  journal = {Information Systems Research},
  volume  = {35},
  number  = {4},
  pages   = {1965--1980},
  year    = {2024},
  doi     = {10.1287/isre.2021.0497}
}

@inproceedings{stein2023,
  author    = {Jake M. L. Stein and Vidminas Vizgirda and Max Van Kleek and Reuben Binns and Jun Zhao and Rui Zhao and Naman Goel and George Chalhoub and Wael S. Albayaydh and Nigel Shadbolt},
  title     = {{`You are you and the app. There's nobody else.'}: Building Worker-Designed Data Institutions within Platform Hegemony},
  booktitle = {Proceedings of the 2023 CHI Conference on Human Factors in Computing Systems},
  year      = {2023},
  doi       = {10.1145/3544548.3581114}
}

@inproceedings{DalalEtAl2023,
  author    = {Dalal, Samantha and Chiem, Ngan and Karbassi, Nikoo and Liu, Yuhan and Monroy-Hern{\'a}ndez, Andr{\'e}s},
  title     = {Understanding Human Intervention in the Platform Economy: A Case Study of an Indie Food Delivery Service},
  booktitle = {Proceedings of the 2023 CHI Conference on Human Factors in Computing Systems},
  series    = {CHI '23},
  year      = {2023},
  publisher = {Association for Computing Machinery},
  address   = {New York, NY, USA},
  articleno = {568},
  pages     = {1--16},
  doi       = {10.1145/3544548.3581517}
}

@article{HuEtAl2024,
  author  = {Hu, Peng and Zeng, Yu and Wang, Dong and Teng, Han},
  title   = {Too Much Light Blinds: The Transparency-Resistance Paradox in Algorithmic Management},
  journal = {Computers in Human Behavior},
  volume  = {161},
  pages   = {108403},
  year    = {2024},
  doi     = {10.1016/j.chb.2024.108403}
}

@article{VertesiEnriquez2025,
  author  = {Vertesi, Janet A. and Enriquez, Diana},
  title   = {The Ghost of Middle Management: Automation, Control, and Heterarchy in the Platform Firm},
  journal = {Sociologica},
  volume  = {19},
  number  = {1},
  pages   = {13--35},
  year    = {2025},
  doi     = {10.6092/issn.1971-8853/16415}
}


\clearpage
\section*{\Large Supplementary Material}

In this document, we provide the detailed participant demographics and interview guides of the paper titled ``When Data Becomes Judgment: Misaligned Interpretations and Accountability in Food Delivery Platforms.''

\begin{table*}[h]
\centering
\caption{Participant Demographics}
\label{table:demographics}
\begin{tabular}{ccccccc}
\toprule
ID & Role & Age & Gender & U/R & Education & Occupation \\
\midrule
U1 & User & 32 & M & R & Bachelor & Professor \\
U2 & User & 27 & M & U & Primary School & Laborer \\
U3 & User & 23 & F & R & Bachelor & Nurse \\
U4 & User & 25 & F & U & Bachelor & Engineer \\
U5 & User & 26 & M & R & Bachelor & Engineer \\
U6 & User & 33 & F & U & Semi-literate & Driver \\
U7 & User & 35 & M & R & Bachelor & Worker \\
U8 & User & 45 & M & U & Bachelor & Designer \\
U9 & User & 55 & F & R & Bachelor & Police Officer \\
U10 & User & 47 & M & U & Primary School & Fisherman \\
U11 & User & 36 & M & R & High School & Manager \\
U12 & User & 31 & F & R & Primary School & Bartender \\
U13 & User & 32 & M & R & High School & Secretary \\
U14 & User & 29 & F & U & High School & Bartender \\
U15 & User & 19 & M & R & High School & Student \\
U16 & User & 22 & F & U & High School & Salesperson \\
U17 & User & 26 & F & R & Master & Consultant \\
U18 & User & 45 & M & R & High School & N/A \\
U19 & User & 44 & M & U & Bachelor & Software Engineer \\
U20 & User & 47 & M & R & High School & Reporter \\
U21 & User & 34 & F & U & Primary School & Farmer \\
U22 & User & 33 & F & U & High School & UX Designer \\
U23 & User & 30 & F & R & Primary School & Researcher \\
\midrule
F1 & Courier & 30 & M & U & Junior High School & Food Delivery Rider \\
F2 & Courier & 32 & M & R & High School & Food Delivery Rider \\
F3 & Courier & 45 & F & U & Bachelor & Food Delivery Rider \\
F4 & Courier & 26 & F & U & High School & Food Delivery Rider \\
F5 & Courier & 29 & F & R & High School & Food Delivery Rider \\
F6 & Courier & 37 & M & U & Junior High School & Food Delivery Rider \\
F7 & Courier & 35 & M & R & Primary School & Food Delivery Rider \\
F8 & Courier & 44 & F & U & Bachelor & Food Delivery Rider \\
F9 & Courier & 40 & M & U & Junior High School & Food Delivery Rider \\
F10 & Courier & 35 & M & U & High School & Food Delivery Rider \\
F11 & Courier & 32 & F & R & Junior High School & Food Delivery Rider \\
F12 & Courier & 32 & M & U & High School & Food Delivery Rider \\
F13 & Courier & 30 & F & U & High School & Food Delivery Rider \\
F14 & Courier & 29 & M & R & Bachelor & Food Delivery Rider \\
F15 & Courier & 36 & M & U & Bachelor & Food Delivery Rider \\
F16 & Courier & 24 & F & U & Bachelor & Food Delivery Rider \\
F17 & Courier & 29 & M & R & High School & Food Delivery Rider \\
\bottomrule
\end{tabular}
\end{table*}

\renewcommand{\arraystretch}{1.2}

\begin{longtable}{p{0.16\textwidth} p{0.34\textwidth} p{0.42\textwidth}}
\caption{Interview Guide for Users} 
\label{table:user_interview}\\
\toprule
\textbf{Theme} & \textbf{Core Question} & \textbf{Possible Probes} \\
\midrule
\endfirsthead

\toprule
\textbf{Theme} & \textbf{Core Question} & \textbf{Possible Probes} \\
\midrule
\endhead

\bottomrule
\endfoot

Overall User Experience & Please think back to your usual food delivery experience. From placing an order to receiving the meal, what is the whole process usually like for you? & Do you usually keep checking your phone? What are you usually looking at? At what moments do you pay special attention to the order? \\

Memorable Experiences & Can you describe one or two food delivery experiences that left a particularly strong impression on you? They may be smooth experiences, or ones that made you feel something was strange, stressful, or confusing. & What happened at that time? Why did this experience leave such a strong impression on you? \\

Moments of Feeling ``Something Was Wrong'' & During the waiting process, are there any moments when you feel that ``something does not seem right'' with the order or that ``something feels off''? & Under what circumstances do you usually get this feeling? What was the first thing you noticed at that moment? \\

How Judgments Are Formed & When you feel that there may be a problem with an order, how do you usually make that judgment? & What kinds of cues do you draw on? Why do those cues lead you to think that way? \\

Understanding Interface Information & When you look at the order interface on the delivery platform, what kinds of information do you usually read from it? & Which kinds of information are most useful to you? Which kinds of information do you feel you still do not quite understand even after seeing them? \\

Interpretation Process & When something on the interface appears unexpected, uncertain, or different from what you anticipated, how do you usually interpret it? & Do you try to guess the reason by yourself? What kinds of explanations do you usually consider? \\

Ways of Responding & When you feel that the delivery process is somewhat unusual, unclear, or not in line with your expectations, what do you usually do? & Do you keep waiting, check repeatedly, contact the courier, contact the restaurant, contact the platform, or do something else? Why? \\

Judging the Courier / System / Restaurant & During the waiting process, if you feel that something has gone wrong at some point, how do you usually judge where the problem may lie? & Who do you tend to think of first? How is that judgment usually formed? \\

Experiences of Mismatch Between Data and Reality & Have you ever had an experience in which you later realized that what you saw at the time was not exactly the same as what had actually happened? & How did you later come to know what really happened? Did this experience affect how you interpret such information afterward? \\

Accumulated Experience Over Time & After using food delivery services for a long time, do you feel that you have gradually developed your own way of making judgments? & How do you think the way you look at orders now differs from when you first started using these services? \\

Feelings About Uncertainty & When you cannot determine what is actually happening, what kinds of feelings does that uncertainty create for you? & Does it make you anxious, annoyed, indifferent, or something else? Does it affect what you do next? \\

Expectations for Future Systems & If you could imagine it, how should future food delivery systems be designed so that you could understand the delivery process more easily? & What would you want the system to tell you more about? In what form would you want it to tell you? \\

Expectations for Records and Judgment Basis & If something really goes wrong during the delivery process, what kinds of information would you want the system to retain to help you understand or judge what happened? & What kinds of records do you think would be more useful? What kinds of information would make you feel that you have a clearer basis for judgment? \\

Imagining an Ideal System & If there were a food delivery system that you considered ``more reasonable,'' what would be the biggest difference between it and current systems? & What would it allow you to see, know, or avoid having to guess? \\

\end{longtable}

\renewcommand{\arraystretch}{1.2}

\begin{longtable}{p{0.16\textwidth} p{0.34\textwidth} p{0.42\textwidth}}
\caption{Interview Guide for Couriers} 
\label{table:courier_interview}\\
\toprule
\textbf{Theme} & \textbf{Core Question} & \textbf{Possible Probes} \\
\midrule
\endfirsthead

\toprule
\textbf{Theme} & \textbf{Core Question} & \textbf{Possible Probes} \\
\midrule
\endhead

\bottomrule
\endfoot

Overall Delivery Experience & Please describe your usual delivery process. From accepting an order to completing the delivery, what is a typical order like for you? & Which parts of the process do you usually pay the most attention to? Which parts are most likely to change? \\

Memorable Delivery Experiences & Can you recall one or two delivery experiences that left a particularly strong impression on you? They may be smooth experiences, or ones that made you feel troubled, misunderstood, or difficult to handle. & What happened at that time? Why did this experience leave such a strong impression on you? \\

Moments of Feeling ``There Might Be a Problem'' & During the delivery process, are there any moments when you feel that ``this order may lead to trouble'' or that ``the customer may become dissatisfied''? & Under what circumstances does this feeling usually arise? What is the first thing you notice? \\

How Judgments Are Formed & When you feel that a particular order may lead to customer dissatisfaction, urging, or later trouble, what do you usually base that judgment on? & Does this judgment come more from experience, platform information, the real-world situation, or something else? \\

Understanding Platform Information & During the delivery process, how do you usually interpret the various kinds of information shown on the platform? & Which kinds of information are most important to you? Which kinds of information do you feel are not fully reliable, or do not fully match the actual situation? \\

Understanding the User's Perspective & Do you think about what the customer may currently be seeing on their phone, and how they may understand your current status? & At what moments do you particularly care about what the customer may be seeing? \\

Experiences of Mismatch Between Data and Reality & Have you ever encountered situations in which what the platform displayed did not quite match what you were actually doing? & What happened at that time? What kinds of effects did this mismatch later bring about? \\

Ways of Responding & When you feel that the situation of an order may cause customers to misunderstand, become anxious, or file a complaint, how do you usually deal with it? & What do you usually do first? Change your route, speed up, contact the customer, or use some other strategy? \\

Decision-Making and Prioritization & In the actual process of handling deliveries, how do you decide what to deal with first, which order to deliver first, or whom to respond to first? & Are there times when your decision is based not only on distance or efficiency, but also on other factors? \\

Communication Experience & Under what circumstances do you usually choose to explain things to the customer or contact them proactively? & In what situations do you feel that explanation is necessary? In what situations do you feel that even if you explain, it may not really help? \\

Experiences of Being Understood / Misunderstood & During the delivery process, have you ever felt that you were understood, or on the contrary, misunderstood? & In what kinds of situations does this feeling usually arise? How did you view it at the time? \\

Accumulated Experience Over Time & After working as a courier for a longer period of time, do you feel that your understanding of the platform, customer reactions, and the overall delivery process has changed? & Compared with when you first started, what is the biggest difference in you now? \\

Expectations for Future Systems & If you could imagine it, how should future food delivery systems be designed so that they could reflect the delivery process more truthfully? & What would you want the system to present more clearly, and what would you want it to omit less often? \\

Expectations for Reducing Misunderstanding & How do you think the system should be designed to reduce misunderstandings between you and customers? & Which explanations would you want the system to provide, rather than requiring you to explain them repeatedly yourself? \\

Imagining an Ideal System & If there were a delivery system that you considered more reasonable, what would be the biggest difference between it and current systems? & How would it help you in your work? How would it make it easier for customers to understand what is actually happening? \\

\end{longtable}


\end{document}